\documentclass[12pt]{article}
\usepackage{graphics,epsfig}

\def\dalemb#1#2{{\vbox{\hrule height .#2pt
        \hbox{\vrule width.#2pt height#1pt \kern#1pt
                \vrule width.#2pt}
        \hrule height.#2pt}}}

\let\a=\alpha \let\b=\beta \let\g=\gamma \let\d=\delta \let\e=\epsilon
\let\z=\zeta  \let\th=\theta  \let\k=\kappa
\let\l=\lambda \let\m=\mu  \let\x=\xi \let\p=\pi %\let\r=\rho
\let\s=\sigma \let\t=\tau    
\let\vp=\varphi \let\vep=\varepsilon
\let\w=\omega       \let\D=\Delta \let\Th=\Theta \let\L=\Lambda
\let\X=\Xi \let\P=\Pi \let\S=\Sigma  \let\Y=\Psi
\let\C=\Chi \let\W=\Omega
\let\la=\label \let\ci=\cite 
  
\def\nn{\nonumber} \def\bd{\begin{document}} \def\ed{\end{document}}
\def\ds{\documentstyle} \let\fr=\frac \let\bl=\bigl \let\br=\bigr
\let\Br=\Bigr \let\Bl=\Bigl
\let\bm=\bibitem
\let\na=\nabla
\def\tU{{\widetilde U}}
\let\pa=\partial \let\ov=\overline
\def\ie{{\it i.e.\ }}
\newcommand{\be}{\begin{equation}}
\newcommand{\ee}{\end{equation}}
\def\ba{\begin{array}}
\def\ea{\end{array}}
\def\ft#1#2{{\textstyle{{\scriptstyle #1}\over {\scriptstyle #2}}}}
\def\fft#1#2{{#1 \over #2}}
\def\F#1#2{{ F_{#1}^{(#2)} }}
\def\cF#1#2{{ {\cal F}_{#1}^{(#2)} }}

\def\={\, =\, }
\def\+{\, +\, }
\def\-{\, -\, }

\def\R{{\bf R}}
\def\sst#1{{\scriptscriptstyle #1}}
\def\oneone{\rlap 1\mkern4mu{\rm l}}
\def\e7{E_{7(+7)}}
\def\td{\tilde}
\def\wtd{\widetilde}
\def\im{{\rm i}}
\newcommand{\ho}[1]{$\, ^{#1}$}
\newcommand{\hoch}[1]{$\, ^{#1}$}
\newcommand{\bea}{\begin{eqnarray}}
\newcommand{\eea}{\end{eqnarray}}
\newcommand{\ra}{\rightarrow}
\newcommand{\lra}{\longrightarrow}
\newcommand{\Lra}{\Leftrightarrow}
\newcommand{\ap}{\alpha^\prime}
\newcommand{\bp}{\tilde \beta^\prime}
\newcommand{\cB}{{\cal B}}
\newcommand{\cO}{{\cal O}}
\newcommand{\vecx}{\vec{x}}
\newcommand{\vecy}{\vec{y}}
\newcommand{\vecp}{\vec{p}}
\newcommand{\vecq}{\vec{q}}
\newcommand{\tr}{{\rm tr} }
\newcommand{\Tr}{{\rm Tr} }

\newcommand{\cL}{{\cal L}}
\newcommand{\cA}{{\cal A}}
\newcommand{\cD}{{\cal D}}
\def\sst#1{{\scriptscriptstyle #1}}
\def\0{{\sst{(0)}}}
\def\1{{\sst{(1)}}}
\def\2{{\sst{(2)}}}
\def\3{{\sst{(3)}}}
\def\4{{\sst{(4)}}}
\def\5{{\sst{(5)}}}
\def\6{{\sst{(6)}}}
\def\7{{\sst{(7)}}}
\def\8{{\sst{(8)}}}
\def\ve{\varepsilon}
\def\vf{\varphi}
\def\F{\Phi}
\def\wg{\wedge}

\def \foot {\footnote}
\def \bi{\bibitem}

\def \tr {{\rm tr}}
\def \ha {{1 \over 2}}
\def \td {\tilde}
\def \ci{\cite}
\def \N {{\mathcal N}}
\def \ww {\Omega}
\def \const {{\rm const}}
\def \ss {\sum_{i=1}^3 }
\def \t {\tau}
\def\S{{\mathcal S} }
\def \nn {\nu}
\def \XX {{\rm X}}

\def \lra {\leftrightarrow}
\def \vom {{\bar \omega}}
\def \E {{\mathcal  E}} \def \J {{\mathcal  J}}
\def \YY {{\rm Y}}

\def \d {\del}
\def \rJ {{J}}
\def \sms {sigma models\ }
\def \sm {sigma model\ }
\def \L {\Lambda}
\def \gl {\ell}
\def \tr {{\rm tr\ }}
\def\z{\zeta}
\def\zi{\zeta_1}
\def\zii{\zeta_2}
\def\K{\mbox{K}}
\def\eE{\mbox{E}}   \def \vt {\vartheta}
\def \vr {\varrho}
\def \wup {w}

\def\dg{\dagger}
\def\a{\alpha}
\def\b{\beta}
\def\e{\varepsilon}
\def\p{\phi}
\def\ap{\alpha^\prime}
\def\I{{\cal I}}

\def\R{{\bf R}}
\def\Z{{\bf Z}}
\def\C{{\bf C}}
\def\P{{\bf P}}
\def\xb{{\bar X}}
\def\Tr{{\rm  Tr}}
\def\tr{{\rm  tr}}

\def \del{\partial}
\def \a {\alpha}
\def \aa {{\a'}}
\def\g{\gamma}
\def\s{\sigma}
\def\z{\zeta}
\def\zi{\zeta_1}
\def\zii{\zeta_2}
\def\ov{\over}

\def\I{{\cal I}}
\def\J{{\mathcal J}}
\def \ok {{1\ov \k}}
\def\LL{{\mathcal L }}
\def \jL {{J}}
\def \om {\omega}
\def \cL {{\mathcal L}} \def \cH {{\mathcal H}}
\def\E{{\mathcal E}}
\def\w{\omega}
\def\b{\beta}
\def\l{\lambda}
\def\eps{\epsilon}
\def\vep{\varepsilon}
\def \De {{\mathcal D}}

\def  \Jt {  {J}_{\rm tot}    }

\def \k {\kappa}
\def\foot{\footnote}
\def \four{{\textstyle {1\ov 4}}}
 \def \third { \textstyle {1\ov 3
}}
\def\det{\hbox{det}}
\def \ci {\cite}

\def \foot {\footnote}
\def \bi{\bibitem}

\def \tr {{\rm tr}}
\def \ha {{1 \over 2}}
\def \tid {\tilde}
\def \vv {{\rm v}}
\def \tl {{\tilde \l}}
\def \XX {{\rm X}}
\def \ta {{\tilde \a}}
\def \fo { {1\ov 4}}
\def \ep {\epsilon}
\def \inti {{\int^{2\pi}_0 {d \sigma \ov 2 \pi}}}

\def \d {\partial}
\def \K {{\rm S}}
\def \el {\ell}
\def \Tr {{\rm Tr}}
\def \P {\Phi}
\def \l  {\lambda}
\def \tl {{\tilde \l}}
\def \bl {{\tilde \l}}
\def \const {{\rm const}}
\def \V {v}

\def \bv {v^*}
\def \vv {{\rm v}}
\def \LL {{\mathcal L}}
\newcommand{\PV}[1]{P_{\!\!_{V_{#1}}}}

\def \bL {\ell}
\def \M {{\mathcal M}}
\def \N {{\mathcal N}}
\def \S {{\rm S}}
\def \vn {\vec n}
\def \tl {\td \l}
\def \td {\tilde}
\def \Prod {\Pi}
\def \O {{\mathcal O}}
\def \Q {{\rm  Q}}
\def \D {\Delta}
\def \N {{\mathcal N}}

\def \m {\mu}
\def \vs {\vec \s}
\def \ie {i.e.}

\def \cD {{\cal D}}

\def  \le  {\l_{\rm eff}}

\def \rS {{\rm S}}
\newcommand{\bra}[1]{\mbox{$\langle #1 |$}}
\newcommand{\ket}[1]{\mbox{$| #1 \rangle$}}

\newcommand{\auth}{AUTHORS}

\def\thb{\bar{\theta}}
\def\Thb{\bar{\Theta}}
\def\barp{\bar{p}}
\def\barq{\bar{q}}
\def\barc{\bar{c}}
\def\bard{\bar{d}}
\def\e{\epsilon}

\def \bi{\bibitem}
\def \la {\label}

\def \l {\lambda}
\def\foot{\footnote}
\def \tl  {{\tilde \l}}
\def \sql {{\sqrt \l}}
\def \adss {$AdS_5 \times S^5$\ }
\newcommand{\rf}[1]{(\ref{#1})}
\def \ov {\over}

\def\th{\theta}
\def\Th{\Theta}
\def\vth{\vartheta}
\def\btheta{{\bar\theta}}
\def\ttheta{{{\tilde\theta}}}
\def\bttheta{{{\bar\ttheta}}}
\def\vth{\vartheta}

\def\ra{\rightarrow}
\def\N{{\cal N}}
\def\F{{\cal F}}
\def\uM{\underline{M}}
\def\uN{\underline{N}}
\def\uP{\underline{P}}
\def\cc{\circ}
\def\eqv{\equiv}

\def\ni{\noindent}
\def \ha{{1\ov 2}}
\def \bw {{\rm w}}

\def\r{{\rm r}}

\def\Y{{\rm Y}}
\def\X{{\rm X}}
\def\tY{\tilde{\rm Y}}
\def\tX{\tilde{\rm X}}
\def\dY{\dot{\rm Y}}
\def\dX{\dot{\rm X}}

\def \J {\mathcal{J}}
\def \del {\partial}

\def\dF{\dot{F}}
\def\dG{\dot{G}}
\def\df{\dot{f}}
\def \E {{\cal E}}
\def \S {{\cal S}}
\def \J {{\cal J}}

\def\ms{\mathcal{S}}
\def\mj{\mathcal{J}}
\def\soj{\fr{\ms}{\mj}}
\def \R {{\bf R}}
\def \om {\omega}
\def \tH {\widetilde H}
\def \bE {\bar E}
\def \x {{\cal X}}

 \def \bb {\bar \beta}
\def \OO {{\cal O}}

\def \hH {\bar H}

\def \w {\omega}

\def \sn {{\rm sn}}
\def \dn {{\rm dn}}

\def \EE {{\rm E}}
\def \KK {{\rm K}}

\begin{document}
\overfullrule=0pt
\parskip=2pt
\parindent=12pt
\headheight=0in \headsep=0in \topmargin=0in \oddsidemargin=0in

\vspace{ -3cm} \thispagestyle{empty}
\vspace{-1cm}
\begin{flushright}
NSF-KITP-05-70; \   \  \
UU-ITP-16/06; \ \ \
CTP-MIT-3678

\end{flushright}

\begin{center}

{\Large\bf
$1/J$
corrections
 to semiclassical   AdS/CFT   states
\\ from   quantum Landau-Lifshitz model

 \vspace{0.01cm}
 }

 \vspace{.5cm} {
 J.A. Minahan$^{a,b,}$\footnote{joseph.minahan@teorfys.uu.se}, A. Tirziu$^{c,}$\footnote{tirziu@mps.ohio-state.edu}
 and A.A.
 Tseytlin$^{c,d,}$\footnote{Also at Imperial College London
 and  Lebedev  Institute, Moscow.
  %tseytlin@mps.ohio-state.edu
 }}\\
 \vskip 0.3cm

{\em $^{a}$Department of Theoretical Physics,\\
Box 803, SE-751 08, Uppsala, Sweden\\
\vskip 0.08cm $^{b}$Center for Theoretical Physics, Massachusetts Institute of Technology\\
Cambridge, MA 02139 USA
\vskip 0.08cm $^{c}$Department of Physics, The Ohio State University,\\
Columbus, OH 43210, USA\\
\vskip 0.08cm $^{d}$ Kavli Institute for Theoretical Physics,  UCSB,  Santa Barbara, CA 93106, USA

     }

\end{center}

\def \W {{\cal E}}

\def \bi{\bibitem}
\def \la {\label}

\def \l {\lambda}
\def\foot{\footnote}
\def \tl  {{\tilde \l}}
\def \sql {{\sqrt \l}}
\def \adss {$AdS_5 \times S^5$\ }

\def \D {\Delta}
\def \thet {\theta}
 \def \t {\tau}
 \def \p {\phi}
 \def \r {\rho}
 \def \rN {{\rm N}}

\def \ov {\over}

\def \varpi {{\rm w}}

% \vspace{0.1cm}

 \begin{abstract}
 %%%%%%%%%%%%%%%%%%%%%%%%%%%%%%%%%
One   way to relate semiclassical
string states   and dual gauge theory states  is to show
the equivalence  between  their low-energy  effective  2d actions.
 The gauge theory effective action, which is represented by an effective
 Landau-Lifshitz (LL) model,  was previously found to match  the string theory
 world-sheet action
 up to  the first two orders  in the effective parameter $\tilde{\lambda}
 ={ \lambda / J^2}$,  where $\lambda$ is the `t Hooft coupling and $J$ is the total $R$-charge.
 Here we address the question if quantizing  the effective   LL action
  reproduces the subleading $1/J$ corrections to  the spin chain energies
 as well as the quantum corrections to the string energies.
  We demonstrate that this is  indeed the case  provided one chooses an appropriate regularization
 of the  effective LL  theory. Expanding near the BPS vacuum, we show that the
 quantum  LL action gives   the same  $1/J$ corrections to energies of BMN
 states   as found previously  on the  gauge theory and string theory sides.  We also compute
  the subleading $1/J^2$ corrections and show that these too match with corrections computed from the
  Bethe ansatz. We also compare the results from the LL
 action with a more direct computation from the spin chain.  We repeat the same
 computation for the $\beta$-deformed  LL action
 and find  that the quantum LL result  is again equal to  the $1/J$
 correction  computed from the  $\beta$-deformed   Bethe ansatz equations.
 We also quantize the LL action near  the rotating circular and
 folded string solutions,   generalizing the   known gauge/string results for $1/J$  corrections to the
 classical  energies.
 We emphasize  the simplicity of this
effective  field theory  approach as compared to the full quantum
string computations.

\end{abstract}
\newpage

\setcounter{equation}{0} \setcounter{footnote}{0}
\setcounter{section}{0}

\renewcommand{\theequation}{1.\arabic{equation}}
 \setcounter{equation}{0}

%%%%%%%%%%%%%%%%%%%%%%%%%%%%%%%%%%%%%%%%%%%%%%%%%%%%%%%%%%%%%%%%
\section{Introduction}
%%%%%%%%%%%%%%%%%%%%%%%%%%%%%%%%%%%%%%%%%%%%%%%%%%%%%%%%%%%%%%%%

Comparing semiclassical string states \ci{bmn,gkp,ft1,ft2}
to ``long'' gauge theory operators \ci{mz,bmsz,bfst} has
turned out to be a  fruitful approach to  exploring AdS/CFT
duality (for reviews and  references  see
\ci{tse1,beis,tse2,zar,swa,plefka}).
A very simple and  clear way of establishing
the correspondence between  ``fast'' strings and
low-energy ``coherent'' spin chain states representing
dual gauge-theory
operators was suggested  in  \ci{kru} and further
clarified and developed  in \ci{krt,mikh,kt}
(for a review  see \ci{tse2}; various extensions were considered in
\ci{hl1,st,mikh1,hl2,castell,castell1,st2}).

In  the simplest nontrivial sector, the $SU(2)$ sector  which has  operators of the form
Tr$(\P^{J_1}_1 \P^{J_2}_2) + ...$,  the corresponding low-energy effective action
 is derived from  the thermodynamic  limit ($J=J_1 + J_2 \gg1 $) of the ferromagnetic spin chain, where the Hamiltonian is the gauge-theory dilatation
operator.    These operators are     dual to strings moving  in the  $R\times S^3$ subspace of
$AdS_5\times S^5$,  where in   the ``fast string'' limit
of the classical  string action one can reduce to the same classical action.
 This Landau-Lifshitz (LL) type
action   serves as  an intuitive  bridge between the
  gauge-theory and
string-theory pictures, suggesting, in particular,   how
a continuous string action and
string picture  may appear  from gauge theory,
as well as suggesting  that quantum string
theory may  have a microscopic spin chain description.

Viewed as an effective low-energy action that emerges
from the two   quantum  ``microscopic'' theories --
the gauge-theory spin chain and the quantum superstring\foot{The limits on the two sides  are, in general, different: (i)
small $\l$, then expansion  in  large $J$, and (ii)
large $\l$ for fixed  $\tl\equiv { \l \ov J^2}$, or large $J$,   and then expansion in $\tl$, see below.} --
the  LL action should not be expected
to lead to a  well-defined  quantum theory.
Yet, supplemented with an appropriate UV cutoff  or regularization
prescription  (as well as with
relevant higher-derivative counterterms)
 it may still be  able to capture    part of the quantum
 corrections to these  ``microscopic''  theories.

 Provided the limits of applicability   of this  quantum LL
 model are understood, it may be  very useful  from both  a
  conceptual  and a technical  point of view.
 On the conceptual side,  the possibility of reproducing certain
 quantum $1/J$ corrections to both string and gauge-theory
 energies from the same   effective LL action
 would  continue to serve as
 an appealing  way  of   understanding   their matching.

 On the technical side, the computation of quantum corrections in the LL
 framework is simpler than the full spin chain
 ({\it e.g.,} Bethe ansatz) computation of finite-size corrections.
 It is also  much simpler than the full superstring computation of
 quantum $\a'$ corrections to energies of string states for the obvious reason that
 here  one  does not  include the contributions of the   bosonic  and fermionic
 modes  which are ``outside'' the given $SU(2)$ sector,
 i.e.  which are absent in the LL  action.
 Omitting these  other string modes  is obviously  not supposed to be correct in general,
 but in some simple cases it may happen  that
  the role of these extra modes may be just
  to provide a particular UV  regularization of the  quantum LL
 result.
 This was first  suggested   in \ci{btz} on the example of the simplest
 circular string solution of \ci{ft2,art} (using results of  earlier work
 in  \ci{ft3,fpt,ptt,lubzar}). Several new   examples
 and non-trivial extensions (like $1/J^2$  corrections to BMN
 energies  or $1/J$ corrections to the folded string energy)
 will be discussed  below.

The quantum LL model may  then provide  a short-cut to
some non-trivial spin chain or  quantum string  results
which would be  much harder to  find    by  more  direct
 computations. This is important since
%This may also lead  to a useful  insight into the structure
%of  quantum string  corrections.
any    new  data about    subleading quantum
corrections to  various semiclassical  string energies
is crucial for  testing all-order conjectures about
the  structure of the quantum \adss string spectrum \ci{afs,s,beiss,szz}.
It is also important to be able to go beyond the leading $1/J$ correction, since the next order
 contains details about the lattice nature of the spin chain.   Being able to compare the gauge theory and string theory results might provide clues to how the lattice nature of the spin chain manifests
 itself in the string theory.

%The hope  is  also to develop a  useful  insight into the structure
%of  quantum string   theory computations
%that may help to understand   when and how
%they  may be   connected with gauge theory computations.

Being an effective field  theory, the quantum LL model should be
supplemented with a regularization prescription, and which
regularization one is to use  should  depend  on the particular
microscopic theory one is trying to approximate.
For example, expanding near the ferromagnetic ground state, which is a BPS  state  of both
gauge theory and string theory,
one needs to assume a normal ordering prescription  for the LL Hamiltonian
so that the vacuum-state energy is not  shifted.
 Choosing a regularization  at higher orders
 is {\it a priori} an open question.
  Still, taking into account  this  regularization ambiguity   by
 introducing free parameters, one  may  able to make non-trivial
 predictions about the dependence  of the energies  on the quantum numbers of
 the fluctuation states.
 %AT
 As we shall discuss below, the normal ordering prescription
appears to be the right one  to match both spin chain and
string theory results  up to  quartic   oscillator order
(and also the right one to match spin chain result in six-order term in the oscillator Hamiltonian).

The
$\zeta$-function regularization is a natural regularization for computing the leading $1/J$
 correction
 to the classical energy
   near a non-trivial  solitonic LL  state
 representing a macroscopic spinning string
 \ci{btz}.\foot{The LL model defined on  $R_t \times S^1_\s$
 has no
 logarithmic UV divergencies, so the $\zeta$-function regularization
   is equivalent to introducing  an explicit cutoff
 $\sum^\infty_{n=1}  e^{-\epsilon n}...$  and dropping  all terms which are
 singular in the limit  $\ep\to 0,$  i.e. is a rather universal  regularization prescription.}
 The utility  of the  $\zeta$-function regularization
 in similar ground-state  energy computations  is well known;
 its use should be justified by additional
 global (space-time) properties that the 2-d theory  should describe.\foot{For example, the use of  $\zeta$-function regularization
 in bosonic $D=26$ string theory in  computing  the ground state energy
 \ci{brink} which
 gives  the standard value for the  tachyon mass
 can be justified  by the requirement of having consistency  between the string
 mass values  and gauge symmetries.
 A similar remark applies to the use of $\zeta$-function  regularization in the
 computation of the  Born-Infeld action in open bosonic string theory
 \ci{FT}.}
 Which regularization to use beyond the first order correction % the quadratic fluctuation
 % order (i.e. in computing the next order $1/J$  corrections to the soliton energy)
 remains  to be understood, but again
  starting with  a solution  depending on  several parameters
  (like winding numbers and spins) one may still get   non-trivial
  predictions   about the quantum corrections
  to  its energy.  Comparing to   other  solutions in various  limits may allow one
   to fix  the required value of the regularization parameters.

   \bigskip
%AT

 Let us now   describe the  contents
   of this  paper.
 In section  2 we first  review
 the derivation of the  $SU(2)$
 LL action    from both gauge theory  (sect. 2.1)
  and  string theory (sect. 2.2),
 emphasizing the limits and  approximations involved.
 We then present the LL action in several equivalent forms
 making explicit its  phase space   structure
 which allows  one to  apply the standard operator  quantization
 procedure.

 In section 3 we  expand the LL action near  its
 trivial vacuum  (corresponding to the ferromagnetic vacuum of the
  spin chain or a point-like  BMN  geodesic  of the  string theory)
 and compute quantum corrections to the   fluctuation
 spectrum.  At  quadratic  fluctuation order we get  the
 leading (order $\tl$)  term in the BMN spectrum (sect. 3.1).
 In sect. 3.2 we consider the quartic fluctuation term in the LL Hamiltonian
 and using  a normal ordering prescription  obtain  the
 leading $1/J$  correction to the $\tl$ term in the BMN spectrum
 which matches the value  from the  gauge theory \ci{mz}  or the
 full superstring computation \ci{parn,callan}.
 In sect. 3.3 we compute  the next order $\tl/J^2$ correction.  At this order there are contributions
 from higher order corrections to the Hamiltonian as well as a second order perturbation theory
 correction coming from the first order correction to the Hamiltonian.  Using $\zeta$-function regularization
 on this latter contribution and a normal ordering prescription on the former, we find agreement with
 results computed from the Bethe ansatz.
 We interpret the  fact that one is able  to reproduce
 the $1/J^2$ spin chain
 result  by quantizing a  continuous 2d  action as an indication that
 the  gauge theory and string theory results (obtained
 in different limits)  may  continue  to match  at $\tl/J^2$ order.
 In sect. 3.4  we extend the computation   to $\tl^2$ order by including
 the 4-derivative (2-loop on the gauge side) term in the LL action.
 The results found from the quantum LL model are compared to the
 order $\tl^2/J $  gauge-theory and string theory results in sect. 3.5
 and complete agreement is found.

In section 4 we compute the Hamiltonian for the quantum fluctuations directly from the spin chain.  Here we find that the Hamiltonian is quartic and automatically normal ordered, and that one can
obtain  the $1/J$ corrections which  are consistent with the Bethe ansatz.    However, we also encounter a subtlety in that the eigenstates of the quadratic piece of the Hamiltonian are not precisely in the  Hilbert space.
Instead, in order to develop  perturbation theory, it is necessary to do the perturbative expansion around
states which are not precisely eigenstates of the quadratic Hamiltonian.    However, we then can perform
a similarity transformation on the Hamiltonian, such that the
transformed  Hamiltonian will have interaction terms of  all orders,
but the states will have the usual Fock space form.  The advantage of building the Hamiltonian this
way is that there are no ambiguities about normal ordering or regularization.

 In section 5 we generalize the LL computation of the
  $\tl/J$  correction to the BMN spectrum to the case of the  $\beta$-deformed
 version of the AdS/CFT \ci{lm,frt1,frt2} (for real deformation parameter $\b$). We demonstrate that the
 corresponding $SU(2)_\b$  anisotropic  (XXZ)
   LL model \ci{frt1}  gives the same
 $1/J$ correction to the analog of the  BMN spectrum   as follows directly from the
 exact spin chain Bethe ansatz  equations of ref. \ci{frt1}.
 This new  non-trivial result for the leading $1/J$ correction in the
  $\beta$-deformed theory  (which was not yet found  directly from the
  corresponding superstring theory)
  may be  used to check  a  consistency  of the corresponding
  ``string Bethe ansatz''  for non-zero deformation parameter
   $\b$  (cf. \ci{afs,frt1,frt2}).

In section 6 we use the quantum LL  model approach to compute
 $1/J$ corrections to the energy of a circular
$J_1=J_2=J/2$  rotating
string solution \ci{ft2,art}   which corresponds
to the  simplest static solitonic  state of the $SU(2)$ LL model.
We first review the result of \ci{btz}  about matching
the $\zeta$-function regularized expression for the leading
1-loop correction to the soliton energy and the corresponding finite-size
correction \ci{btz,lopez} to the thermodynamic-limit
 spin chain result, which is also equal to the  leading term in the exact
 1-loop string-theory expression found in \ci{fpt,ptt}.
In sect. 6.2 we extend the computation to the  next sub-subleading
$\tl/J$ order  using   second order quantum-mechanical
 perturbation
theory for the LL Hamiltonian  (the $\tl/J$  correction was not yet computed from  either the
gauge-theory  spin chain or the string).
    We   point out the existence of the
regularization ambiguity which remains to be fixed: it is no longer clear
that the $\zeta$-function regularization should continue to correspond to
either of the two microscopic theories --  spin chain or superstring.
In sect. 6.3  we include  the  ``gauge-theory 2-loop'' $\tl^2$ term in the classical LL
action  and  again compute the leading 1-loop correction to
 the classical  energy. The result matches the second order term in the
 formal expansion  of the exact finite one-loop string correction to
 the spinning string energy \ci{fpt,ptt},  provided one uses  the
 $\zeta$-function regularization to define the formal expression for
 the   string-theory coefficient (this prescription
 is the one  consistent to the given $\l^2$ order
  with the Bethe ansatz  for a similar $SL(2)$ case \ci{szz,bt}).

In section 7 we  attempt to repeat what was done in  section 6 in
the case of  a more complicated  solitonic  LL solution
representing a folded  2-spin ($J_1,J_2) $ string \ci{ft4,bmsz} rotating  in
$S^5$.  Here  the  LL fluctuation Lagrangian explicitly depends on
(elliptic functions of) the spatial coordinate
$\s$,  and  computing  its
spectrum exactly  appears difficult. Instead, we use  the  ``short string'' expansion  in the
parameter $\a= {J_2/J}$\  ($J=J_1 + J_2$)
 and compute the two coefficients in the small
$\a$ expansion of the 1-loop correction to the folded string
energy. The corresponding results  for both the string or  the spin chain
remain to be obtained, and  we expect them to match the
result of the quantum LL computation.

%AT
In Appendix A we compute the energy of  $M$-impurity  near-BMN state up to $1/J^2$ order
in the $SU(2)$ sector
 directly from  the  Bethe ansatz.
In Appendix B  we give some technical details for the  evaluation of  sums in
sect. 6.2. Appendix C  contains  a computation of the numerical coefficient
of  the $\a^2$ term in the $1/J$ correction to the folded string energy.
In Appendix D we present the results of the similar
computation for the $1/J$ correction to the energy of $(S,J)$ folded
string in the $SL(2)$ sector.

\bigskip

There are a number of obvious open problems,
including the range of applicability of  the quantum LL model and the
choice of regularization.
There are several
 computations  similar to the ones described in this paper
 that would be useful to carry out.
It would be interesting to repeat the  computation of the $1/J^2$
correction to  the energy of circular  string  in section 5 for a
similar circular  $(S,J)$  solution \ci{art} in the $SL(2)$
sector. This latter solution is stable and thus the result could
be consistently  compared to the Bethe ansatz  one for  the
sub-subleading correction  which should follow from a
generalization of  the analysis of the $1/J$ correction in
\ci{btz,lopez,szz}.

 It  is possible  to  repeat similar
computations in the  $SU(3)$ sector, and compare the results with
string theory \cite{callan} and gauge theory \cite{f}.
%AT
Another sector to  consider is the $SU(1|1)$ sector  \ci{st2,aaf}
where results for $1/J$ corrections should  be easier to obtain.
One could
also find the quantum LL corrections to the
  fluctuations  near the non-trivial $(J,J,J)$  vacuum  of
   the  $SU(3)_{\b}$  sector of the $\b$-deformed
version of AdS/CFT  \ci{frt2}.
Another computation worth doing is
for the $\tilde{\lambda}^2/J^2$ corrections to BMN states. This
can be done on one hand by using quantum LL, and on the other hand
by using Bethe ansatz. The two computations are expected to agree,
as one expects full agreement between the gauge and  the string theory up
to two loops (i.e. at orders $\l$ and $\lambda^2$).

\renewcommand{\theequation}{2.\arabic{equation}}
 \setcounter{equation}{0}

%%%%%%%%%%%%%%%%%%%%%%%%%%%%%%%%%%%%%%%%%%%%%%%%%%%%%%%
\section{Landau-Lifshitz action in the $SU(2)$ sector}
%%%%%%%%%%%%%%%%%%%%%%%%%%%%%%%%%%%%%%%%%%%%%%%%%%%%%%%

Let us  start  with recalling the  derivation of the
effective Landau-Lifshitz action on  both  the gauge theory
(spin chain)
  and the string theory sides \ci{kru,krt,tse2}.

%%%%%%%%%%%%%%%%%%%%%%%%%%%%%%%%%%%%%%%%%%%%%%%%%%%%%%%%%%%%
  \subsection{ LL action from  gauge theory  }
%%%%%%%%%%%%%%%%%%%%%%%%%%%%%%%%%%%%%%%%%%%%%%%%%%%%%%%

The planar 1-loop dilatation operator  of
the  $\N=4$ SYM theory
coincides  with the
   Hamiltonian of the ferromagnetic Heisenberg  XXX$_{1/2}$ model
   \ci{mz}  ($\l= g^2_{\rm YM} N$)
\be \la{ferr}
H=  { \l \ov (4 \pi)^2} \sum^J_{l=1}(I - {\vs}_l \cdot { \vs}_{l+1})
 \ . \ee
To describe a subsector of eigenstates  that correspond to
``semiclassical'' low-energy part of the  spectrum  it is useful to use
the coherent states which are products of   spin coherent states
at each site with the characteristic property
$\bra{\vec n  } \vec \s  \ket{\vec n} =  \vec n$, \ $\vec n^2 =1$.
In general, one can rewrite the usual phase space path integral
as an integral over the overcomplete set of coherent states:
\be\la{zz}
 Z= \int [dU] \ e^{i \S[U]}  \ , \ \ \ \ \ \ \ \ \
\S= \int dt \bigg( \bra{U }  i  { d \ov dt} \ket{U}  -
\bra{U }  H  \ket{U} \bigg) \ . \ee
 The first (``Wess-Zumino'')
term in the action $\sim i U^* { d \ov dt} {U}$
is the analog of the usual $ p \dot q$ term
 in the phase-space action. Applying this to the case of the
Heisenberg
spin chain Hamiltonian
\rf{ferr} one ends up with  with the following action
for the coherent state variables $\vec n_l(t)$ at sites $l=1,...,J$
($U^\dagger \vec \s U = \vec n $):
\be \S=\int dt \sum^J_{l=1}  \bigg[ \vec C (n_l) \cdot  \del_t  \vec n_l
- { \l \ov 2 (4 \pi)^2 }  ( \vec n_{l+1}
- \vec n_l)^2  \bigg] \ . \ee
Here $d C= \epsilon^{ijk} n_i d n_j \wedge d n_k$,
i.e. $\vec C$ is a monopole potential on $S^2$.
In local coordinates (at  each site $l$) one has
$\vec n = (\sin 2 \psi  \ \cos 2 \vp, \
 \sin 2 \psi  \ \sin 2 \vp,\ \cos 2 \psi )$,
\ $\vec  C \cdot  d \vec n =  \cos 2 \psi\   d \vp$.
So far,  no approximation was made.
If we  now consider the large $J$ limit and concentrate
on low-energy  excitations of the spin chain
then  $n_i$ should change slowly
from site to site and it is natural  to take the continuum limit
by introducing the 2-d field $ \vec n(t,\s)= \{ \vec n (t,
 { 2 \pi \ov J} l) \}$, $l=1,...,J$.
  Then the action becomes ($\del_1 =\del_\s$)
\be \la{con}
\S= J \int dt \inti  \left[ \vec C \cdot  \del_t  \vec n -
{1 \ov 8} \tl (\del_1 \vec n)^2  + ...
%O( { 1 \ov J^2}  (\del_1^2 n)^2)
\right] \ , \ \ \ \ \     \tl \equiv {\l \ov J^2} \ ,    \ee
where dots stand for higher derivative terms suppressed by
$1 \ov J$. The leading correction scales  as
$ { 1 \ov J^2}  (\del_1^2 n)^2$. Indeed,
\begin{equation}
\vec n_{l+1}-\vec  n_{l}=\frac{2\pi}{J}
\partial_{1}\vec n+\frac{1}{2}\left(\frac{2\pi}{J}\right)^2
\partial_{1}^2\vec  n+\frac{1}{6} \left(\frac{2\pi}{J}\right)^3 \partial_{1}^3
\vec n+...\ , \la{iio}
\end{equation}
i.e.
\begin{equation}
\frac{\lambda}{2(4\pi)^2}\sum_{l=1}^{J}(\vec  n_{l+1}-\vec
n_{l})^2\rightarrow\frac{\lambda}{J}
\left[ (\partial_{1}\vec   n )^2-\frac{\pi^2}{3
J^2}(\partial_{1}^{2}\vec   n)^2+...\right]\  . \la{ii}
\end{equation}
Observing that $J$ appears  in front of the action and thus plays
the role of the  inverse Planck constant, we may expect
that the classical  Landau-Lifshitz (LL)  action \rf{con}  with the
equations of motion
\be  \del_t  n_i = \ha \tl \epsilon_{ijk} n_j \del^2_1 n_k \ee
should be
 describing the low-energy part of the spectrum with energies scaling
 as $ J \tl$  to leading  order in the quantum $1/J$ expansion.
 Since the first subleading term  in \rf{ii} scales
 as $ 1/J^2$  one may expect  that  order  $1/J$ corrections
 to the  energies  of the corresponding low-energy states
 can be found by quantizing just the continuous
 LL action.
 However, to capture $1/J^2$ and higher
  order corrections  to the  energies as described by the discrete
  Heisenberg Hamiltonian
  one
 would need to add higher-derivative  terms  omitted in taking the
 continuum limit.
 This will be discussed in detail below.

  As we shall see,
 extending the  observation in \ci{btz},
 the semiclassical quantization of the LL action does allow one to
 reproduce the $1/J$  corrections (as
  found, e.g.,  from the discrete Bethe ansatz)
  in a very simple  way provided one uses
  an appropriate  UV regularization. As usual in an  effective field
  theory approach, the underlying microscopic UV finite
   theory (spin chain)  dictates a particular choice of a
   regularization. There is no a priori choice of this regularization
   within the continuous effective theory, unless one uses some
   additional conditions like that energies  of some BPS states
   should not be changed by  $1/J$ corrections. We shall  provide
 examples  of this in what follows.

 %%%%%%%%%%%%%%%%%%%%%%%%%%%%%%%%%%%%%%%%%%%%%%%%%%%%%S
 \subsection{ LL action from string theory}
%%%%%%%%%%%%%%%%%%%%%%%%%%%%%%%%%%%%%%%%%%%%%%%%%%%%%%%%%

  The same  LL action  appears \ci{kru,krt}
   as an effective action
 on the string theory side too  where one
 also considers a (different)  semiclassical  limit.
 One  concentrates on  a sector of states for which large $J$ expansion is
 equivalent to  quantum string  (inverse string tension)
 expansion. One  first takes $\l$ large,  or, equivalently (for given sector of states),
 $J$ large to suppress quantum corrections
  and then expands the classical string action  in the
  inverse of the effective  semiclassical parameter
  $\tl \equiv {\l \ov J^2}$.  The derivation goes through  the following steps \ci{kru,krt,kt}:
  (i) one isolates a  ``fast'' coordinate $\a$ whose momentum
   $p_\a$ is
large for given  class of string configurations;
(ii) one gauge-fixes $t = \tau$ and
$ p_\a = J$ (or $\td \a = J \s$ where $\td \a $ is ``T-dual''
to $\a$);
(iii) one expands the action in  derivatives of
``slow''  coordinates, or equivalently,
in $\sqrt {\tl}= { 1 \ov \J}$.
In
  the $SU(2)$ sector of string states
carrying two large spins in $S^5$, with string motions restricted
to $S^3$ part of $S^5$,
 the relevant part of the \adss metric is
$ds^2= - dt^2 + d\XX_i d\XX_i^*$, with $\XX_i \XX_i^*=1$.
Setting
\be  \XX_1 = X_1 + i X_2 = U_1 e^{i \a}\ , \ \ \ \ \ \ \ \ \
\XX_2 = X_3 + i X_4 = U_2 e^{i \a}\ , \ \ \ \ \ \ \ \
 U_a U^*_a=1\ , \ee
we identify  $\a$ as  a  coordinate  associated to the
total  spin in the two planes  and
$U_i$   as  ``slow''
coordinates determining the
 ``transverse'' string profile.
  Then
\be  d\XX_a d\XX_a^* = ( d \a + C)^2 + D U_a DU^*_a   , \ \ \ \ \ \
C= - i U^*_a dU_a , \ \ \ \ \ \ \ DU_a = d U_a -i C U_a  .  \ee
Introducing $\vec n = U^\dagger \vec \s U, \ U=(U_1,U_2)$
we get
\be
d\XX_a d\XX_a^* = (D \a)^2 +  { 1 \ov 4} ( d \vec n)^2 \ ,
\ \ \ \ \ \ \ \ \ \ \
\ D\a= d \a + C(n) \ .\ee
 Writing the resulting string sigma model action in phase space form,
 one may fix the
gauge $t= \tau, \ p_\a =$const$=J$.
Making the key assumption that the evolution of $U_a$
in $t$ is slow, i.e. the time derivatives are suppressed
(which can be implemented by  rescaling $t$ by $\tl$ and
expanding in powers of $\tl$),
we find, to the  leading order in $\tl$,
\be \la{coon}
S=J\int dt \int_{0}^{2\pi}\frac{d\sigma}{2\pi}\ L
\ , \ \ \ \ \ \ \
L = - i U^*_a \del_t U_a - \ha \tl |D_1 U_a|^2  + O(\tl^2) \ . \ee
This  becomes   the same as the $CP^1$ Landau-Lifshitz
action
 \rf{con} when written in terms of
$\vec n$.
The agreement between the low-energy effective actions
 on the spin
chain and one the string side implies
the matching of
energies of  the coherent states representing
 configurations with two
large spins (and also the matching of near-by fluctuations).

This  agreement between the effective LL actions
extends also  to the next $\tl^2$ order \ci{krt}.
To get the $\tl^2$ term in \rf{coon} one is to do a field redefinition
to trade  time derivatives for spatial ones; the result is
a generalization of \rf{con}
\begin{equation}
L=\vec C  \cdot    \del_t \vec n
-\frac{\tilde{\lambda}}{8}(\partial_{1}n_{i})^2+\frac{\tilde{\lambda}^2}{32}
\left[(\partial_{1}^{2}n_{i})^2-\frac{3}{4}(\partial_{1}n_{i})^4\right]
+O(\tilde{\lambda}^3)\ .
\label{2loop}
\end{equation}
The same  action is found on the spin chain side by
adding to the   dilatation operator \rf{ferr}  the 2-loop
 term  \ci{bks},
 $H_2=  { \l^2 \ov (4 \pi)^4} \sum^J_{l=1}
 (-3 +4 {\vs}_l \cdot { \vs}_{l+1}  -
   {\vs}_l \cdot { \vs}_{l+2})$,
 taking coherent state expectation value and also
 including a quantum correction \ci{krt}.

This agreement   between the effective  actions
is  rather remarkable, given
that the limits taken on the two sides  of the duality
are different \ci{ss,bds}:  on gauge theory  side  we first take
$\l$ small and then expand in large  $J$ isolating  contributions
depending on $\tl = {\l \ov J^2}$,    while on string side we first take
$J$ large  with $\tl =$fixed
to suppress quantum corrections and then expanded in $\tl$.

 A natural question is if  this  matching  continues
 at  subleading $1/J$ order, i.e. if corrections to
 thermodynamic limit  on the 1-loop
 spin chain side are the same as  the leading 1-loop
 corrections on the  string theory side to the same linear order in
 $\tl$. This matching was found  on several  explicit examples:
 near BPS (BMN) states  \ci{parn,callan} and circular strings \ci{fpt,ptt,btz}
 (see also \ci{lubzar,lopez,bfrey}).
 A simple  way to understand why this happens   was suggested in
 \ci{btz} by computing the leading quantum correction to the energy of
 circular string state
  directly at the level of the effective LL model.
  From spin chain perspective,  quantizing LL action  should indeed
  correctly capture
  the leading $1/J$ correction to the energy,
   provided one uses an appropriate
  regularization  equivalent to  the one  built into the discrete spin chain
  (Bethe ansatz) computation.

  On the  string theory side, the full 1-loop correction to the energy
  \ci{ft3,fpt,ptt}  contains the contribution of not only the
  2 ``transverse''  fluctuations described by the LL action but also
 2 other $S^5$ fluctuations outside of $S^3$,
  4 $AdS_5$  fluctuations and also of  the fermionic fluctuations
  that are crucial for  finiteness of the result.
  Remarkably, it  was observed in \ci{btz}  that the leading $1/J$
  (order $\tl$)
   contribution of
  ``external'' bosonic and  fermionic fluctuations  has a
   trivial   ``counterterm''-type  form,
    i.e. the full string result  can  be correctly  reproduced  by
    quantizing only the two ``internal'' LL  fluctuations
    and using a specific ($\zeta$-function) regularization.

Thus, as on the spin chain  side, the usual  effective field theory
ideology seems to apply: the full string theory
can be  interpreted as a  UV finite microscopic theory   which
contains  (when
expanded near a particular circular string background with $\J$ large)
  ``light''  and  ``heavy''   fluctuation modes,
with the ``light'' modes described by the effective
  LL  action, and the role of the ``heavy''  modes
  being to provide a  regularization prescription
  for the  quantum effective field theory.
 While the  two microscopic  theories -- the  spin chain and the string theory --
 are very different, both lead  to the same LL  action in the classical
 limit, and, moreover, to the same quantum version of it
 with  the same regularization prescription.

Below we would like to explore other examples  when    this matching of the
quantum corrections  continues to happen.
One motivation  for starting directly with a quantum LL  Hamiltonian
 is technical simplicity:  both spin chain and full string theory
 computations of subleading corrections are rather involved, while the quantum LL
   framework provides  simple  framework
   for  model computations and  checking conjectures
   about structure of quantum corrections.

 To prepare  for the discussion of
 particular cases   let us first present  the explicit
 form of the LL Hamiltonian  and its quantization.

%%%%%%%%%%%%%%%%%%%%%%%%%%%%%%%%%%%%%%%%%%%%%%%%%%%%%S
 \subsection{Canonical structure of the LL Lagrangian }
%%%%%%%%%%%%%%%%%%%%%%%%%%%%%%%%%%%%%%%%%%%%%%%%%%%%%%%%%

Let us start  with rewriting the LL
Lagrangian \rf{con} or \rf{coon}  in terms of two independent
fields.
Solving the constraint $n_i n_i =1$  as
 $n_3=\sqrt{1-n_{1}^2-n_{2}^2}$  we get the following  $SO(2)$ invariant  expression for the
 Lagrangian in terms of $n_1$ and $n_2$   ($a,b=1,2$; \  $n^2= n_a n_a$)
\begin{equation}\la{laag}
L=  h^2(n)
  \epsilon_{ab}\dot{n}_{a}n_{b} - H (n) \ , \ee
  \be \la{fag}
 h^2(n)=\frac{1-\sqrt{1-n^2}}{2n^2}=
 {1 \ov 4}   + { 1 \ov 16} n^2 + {1\ov 32} n^4
 + ... \ , \ee
 \be \la{haag}
  H(n)=  \frac{\tilde{\lambda}}{8} \left[  n'^2_{a}+
\frac{1}{1-n^2}\ (n_{a}n'_{a})^2 \right] \label{vector}\ ,
\end{equation}
where we use dot and prime for time and space derivatives.
We have added and subtracted  a  total derivative term
$\frac{\epsilon_{ab}\dot{n}_{a}n_{b}}{2n_{a}^2}=\frac{1}{2}
\frac{\partial}{\partial t}\left(\arctan
\frac{n_{1}}{n_{2}}\right)$
to make the function $h$  have regular expansion near $n_a=0$.
Thus \rf{laag}
 may be  interpreted as a phase-space  Lagrangian
 with, say, $n_1$  being a coordinate  and $n_2$
 related to  its momentum.
%  (note that the Poisson
%structure  is ultra-local, i.e. there are no derivatives in
%$\s$).

In what follows  we shall expand the LL
action near particular   solutions and quantize.
To simplify the quantization it is useful
 to put $L$ into the standard canonical form.
This   can be done  by  the field redefinition $n_a \to z_a $
(which is regular at the origin)
\begin{equation}
z_{a}=h(n)\ n_{a} \ , \ \ \ \ \ \ \ \ \ \ n_{a}= 2\sqrt{1-z^2}\
z_{a} \ . \la{nz} \ee Then we get ($z^2= z_a z_a$)
\begin{equation}
L=\epsilon_{ab}\dot{z}_{a}z_{b}  -H(z) \ , \ \ \ \ \ \ \ \ H(z)
=
 \frac{\tilde{\lambda}}{2}\left[(1-z^2)\ z_{a}'^2\ +\ \frac{2-z^2}{1-z^2}\ (z_{a}z_{a}')^2\right] .  \label{vv}
\end{equation}
Note that the LL  Hamiltonian $H(n)$  or $H(z)$ is the same as for a sigma model
on a sphere $S^{2}$ written in different coordinates.

Having the Lagrangian in the
standard $L=p \dot q - H(p,q)$ form the quantization is
straightforward: we are to promote $z_a$ to operators, impose the
canonical commutation relation $ [z_1(t,\s),z_2 (t,\s')]=  i\pi
\delta (\s - \s') $ and decide how to define  the  quantum
Hamiltonian $H(z)$, i.e. how to order the ``coordinate'' and
``momentum'' operators in it. We will discuss this on explicit
examples below.

Let us  mention also  another explicit parametrization  of the
LL Lagrangian in terms of angles $\psi,\vp$.\foot{In terms
of global angular coordinates of $S^5$ with the metric
$ds^2=dt^2+d\gamma^2+\cos^{2} \gamma\ d\varphi_{3}^2+\sin^{2}
\gamma\ (d\psi^2+\cos^{2}\psi\ d\varphi_{1}^2+\sin^{2}\psi\
d\varphi_{2}^2) $ we have
$\varphi=\frac{\varphi_{1}-\varphi_{2}}{2}$, and
$\alpha=\frac{\varphi_{1}+\varphi_{2}}{2}$.}
 If we set
\begin{equation}
U_{1}=\cos \psi\ e^{i\varphi}, \quad U_{2}=\sin \psi\ e^{-i\varphi}\ , \ \
\ \ \ \ \ \
\vec{n}=(\sin 2\psi
\cos 2\varphi,\sin 2\psi \sin 2\varphi, \cos 2\psi) \ ,
\label{angle}
\end{equation}
then
\begin{equation}
L=\cos 2\psi\
\dot{\varphi}-\frac{\tilde{\lambda}}{2}\left(\psi'^2+
\sin^2 2\psi\ \varphi'^2 \right)  \ .    \label{LLu}
\end{equation}
%Let us consider fluctuations around a particular solution. With
Setting
 \be  \xi = \ha  \cos 2\psi \ee
 we get
\begin{equation}
L=2\xi
\dot{\varphi}-\frac{\tilde{\lambda}}{2}\left[\frac{\xi'^2}{1-4\xi^2}+
(1-4\xi^2) \varphi'^2 \right] \ .
\label{fluct4}
\end{equation}
%This form of the Lagrangian is simpler than the one obtained above
%using unit vector coordinates. One can see that we can expand this
This form of the LL Lagrangian is useful for  expansion
around any particular solution  with $\psi\not=0$;
near the  solution  with $\psi=0$ or
$1- 4 \xi^2=0$   we get
the usual  polar-angle type of singularity
and  should   use instead  the  ``cartesian''
form of $L$ in  (\ref{vv})
which is regular at the origin, i.e. near $n_a=0$.

%%%%%%%%%%%%%%%%%%%%%new%%%%%%%%%%%%%%%%%%%%%%%%%%%%%%%%%%%%%%%%%%

\renewcommand{\theequation}{3.\arabic{equation}}
 \setcounter{equation}{0}
%%%%%%%%%%%%%%%%%%%%%%%%%%%%%%%%%%%%%%%%%%%%%%%%
\section{Quantization near the BPS vacuum:\\
$1/J$ and $1/J^2$   corrections to BMN spectrum}
%%%%%%%%%%%%%%%%%%%%%%%%%%%%%%%%%%%%%%%%%%%%%%

\subsection{Generalities and BMN  spectrum }

Let us now try to reproduce $1/J$ corrections to
the leading terms in the  BMN spectrum of fluctuations near
 the vacuum solution
\begin{equation}
\XX_{1}=e^{i \mathcal{J}t}, \quad \XX_{2}=0 \label{geo}\ ,
\ \ \ \ \ \psi=0\ , \ \ \ \   \varphi=0  \ ,    \ \ \ {\rm
i.e.}
\ \ \ \vec n=(0,0,1)\ , \ \ \
\end{equation}
by quantizing the above LL action.
These corrections can be found from the  Bethe ansatz on the spin chain
\ci{mz,beisert} or from direct superstring quantization
\ci{parn,callan}, but the derivation  from the  LL action
turns out to be  much simpler.

Expanding  near this
vacuum corresponds to expansion near $n_a=0$ in \rf{laag}
 or $z_a=0$ in
\rf{vv}. Observing that  the factor $J$ in
front of  the LL action \rf{con},\rf{coon} plays the role of
the inverse
Planck constant, it is natural to rescale  $z_a$ as
\begin{equation}
z_{1}= \frac{1}{\sqrt{J}}\ f\ ,\ \ \ \  \quad z_{2}=
\frac{1}{\sqrt{J}}\ g\ , \la{fg}
\end{equation}
so that powers of $1/J$ will play the role of
coupling constants in the  non-linear  LL Hamiltonian for the fluctuations.
 To  sixth  order in the fluctuation
 fields $f,g$  we get
\begin{equation}
S=\int dt \int_{0}^{2\pi}\frac{d\sigma}{2\pi}\ ( 2\dot{f}g- H )
\ , \  \ \ \ \
H = H_2 + H_4 +H_6 + ... \ , \la{qua} \ee
\be H_2
= \ha \tl (f'^2+g'^2)\ ,  \label{quad}  \ee
\be
 H_4=  { \tl  \ov 2 J } \left[
2(f f'+g g')^2-(f^2+g^2)(f'^2+g'^2)\right]  \ , \label{quartic}
\end{equation}
\be
 H_6=
   { \tl  \ov 2 J^2  }  (f^2+g^2)(ff'+gg')^2  \ .    \label{six}
\end{equation}
Let us first consider the quadratic approximation.
The linearized equations of motion for fluctuations are
\begin{equation}
\dot{f}=-\frac{\tilde{\lambda}}{2}g'', \quad\ \ \ \ \ \ \
\dot{g}=\frac{\tilde{\lambda}}{2}f''\ ,  \label{eqmotion}
\end{equation}
and their  solution may be written as  ($f,g$ are real)
\begin{equation}\label{fsol}
f(t,\sigma)=\frac{1}{2}\sum_{n=-\infty}^{\infty}(a_{n}e^{-i\omega_{n}t+in\sigma}+
a_{n}^{\dagger}e^{i\omega_{n}t-in\sigma})\ , \ \ \ \ \ \
\omega_{n} =  \ha \tl n^2  \ , \la{lo}
\end{equation}
\begin{equation}\label{gsol}
g(t,\sigma)=\frac{1}{2}\sum_{n=-\infty}^{\infty}(-ia_{n}e^{-i\omega_{n}t+in\sigma}+i
a_{n}^{\dagger}e^{i\omega_{n}t-in\sigma})\ . \la{loo}
\end{equation}
For  each solution of LL equations of motion one needs
also to impose an
extra constraint that the total momentum in $\sigma$-direction  is
zero \cite{kru,kt}
\begin{equation}
P=-i\int_{0}^{2\pi}\frac{d\sigma}{2\pi}U_{a}^{*}U'_{a}=
\int_{0}^{2\pi}\frac{d\sigma}{2\pi}  \cos 2\psi\ \varphi' =0 \ .
\end{equation}
Expanding  near the vacuum, we obtain the
constraint on fluctuations
\begin{equation} \la{hou}
P =2\int_{0}^{2\pi} \frac{d\sigma}{2\pi} f'
g=-\sum_{n=-\infty}^{\infty}n a_{n}^{*}a_{n}=0\ .
\end{equation}
Upon quantization  \rf{eqmotion} become
the equations of motion for the operators $f,g$
\begin{equation}\la{op}
\dot{f}=i[\bar H_2,f], \quad \quad \dot{g}=i[\bar H_2,g] \ , \ \ \ \ \ \
\bar H_2 \equiv   \int_{0}^{2\pi} \frac{d\sigma}{2\pi}\  H_2 \ ,
\end{equation}
provided we use the  canonical commutation relations
\begin{equation}\la{opi}
[f(t,\sigma),f(t,\sigma')]=0\  , \quad
[g(t,\sigma),g(t,\sigma')]=0\ , \ \ \ \
[f(t,\sigma),g(t,\sigma')]=i\pi \delta(\sigma-\sigma')\ .
\end{equation}
Then  the coefficients  in \rf{lo},\rf{loo}   satisfy
\begin{equation}\la{ui}
[a_{n},a_{m}^{\dagger}]=\delta_{n-m} \ ,
\end{equation}
i.e.   $a_{n}$, $a_{n}^{\dagger}$
can be interpreted as  annihilation and creation
operators,
% We denote the
%state $|...,\textbf{N}_{n},...>$, the state in which the
%oscillator of type $n$ is excited on the level $N_{n}$
%\begin{equation}
%a_{n}^{\dagger}a_{n}|...,\textbf{N}_{n},...>=N_{n}|...,\textbf{N}_{n},...>
%\end{equation}
with the  vacuum state $|0\rangle$   defined  by $a_n |0\rangle=0, \ n=0,\pm 1, ...$.
 A general oscillator state is
\begin{equation}
|\Psi>=\prod_{n=-\infty}^{\infty}
\frac{(a_{n}^{\dagger})^{k_{n}}}{\sqrt{k_{n}!}}|0\rangle \ . \label{states}
\end{equation}
The  integrated  Hamiltonian $\bar H_2$  becomes
\begin{equation}\label{ham2}
\bar H_2  =
\frac{\tilde{\lambda}}{4}\sum_{n=-\infty}^{\infty}n^2(a_{n}a_{n}^{\dagger}+a_{n}^{\dagger}
a_{n})=\frac{\tilde{\lambda}}{2}\sum_{n=-\infty}^{\infty}n^2
a_{n}^{\dagger} a_{n} + e_0 \  ,\ee \be
 e_0= \frac{\tilde{\lambda}}{2} \sum_{n=-\infty}^{\infty}n^{2}\ .
\end{equation}
At this point we should
add the  requirement
 that the vacuum  energy should be zero:
 $$ e_0 =0 \ . $$
  We know that the  BMN vacuum is
 a BPS state in both gauge theory and string theory.
 This  amounts  to normal ordering prescription  for the quadratic
 Hamiltonian  or use of a regularization (e.g., the $\zeta$-function one)
 in which $e_0$ is set to zero.  We stress  that this
 condition is an additional constraint  one needs to  impose to
 make quantum LL theory consistent with   ``microscopic''
 spin chain or string theory. Similar conditions
 will be needed  at higher  orders to fix the
 regularization ambiguity present in quantum LL theory.

% This procedure of dropping
%all terms due to normal ordering is equivalent with asking that
%the Hamiltonian annihilates the vacuum and $1$-particle states
%$a_{n}|0\rangle$. The logic for such a requirement is that on the gauge
%theory side the $1$-impurity state corresponds to the BPS operator
%$Tr(Z^{J}\Phi)$. As we go to higher orders in fluctuations and
%$\tilde{\lambda}$, we use the prescription of zeta-function
%regularization and dropping all terms that do not annihilate the
%vacuum and $1$-particle states. We will see that this prescription
%and normal ordering are identical at quadratic and quartic order,
%but different at sixth order in fluctuations.

The momentum condition \rf{hou}
becomes the  constraint on physical states:
\be \la{cons}
\sum_{n=-\infty}^{\infty}n a_{n}^{\dagger}a_{n}|\Psi\rangle  =0\ , \ \ \ \ \ \
\sum_{n=-\infty}^{\infty}n k_n =0 \ . \ee
Let us  consider
 $M$-impurity states as oscillator states with $k_n=1$:
 \begin{equation}
|M\rangle
%\equiv|\textbf{1}_{n_{1}}...\textbf{1}_{n_{M}}>
=
a_{n_{1}}^{\dagger}...a_{n_{M}}^{\dagger}|0\rangle \ .
\end{equation}
 For simplicity we  shall consider states  with all $n_j$  being
 different;
 computations for more general states
 with several equal $n_j$   are similar, at least for first order corrections in $1/J$.

The zero-momentum condition \rf{cons}   gives
\begin{equation}\label{momcond}
\sum_{j=1}^{M}n_{j}=0 \ ,
\end{equation}
%which is the condition we need to impose on physical states. Note
which is also the
 condition on BMN states  present in both string and gauge theory.

The leading  term in the energy of an $M$-impurity state is then
\begin{equation}
\langle M|\bar H_2 |M\rangle=\frac{\tilde{\lambda}}{2}\sum_{j=1}^{M}n_{j}^{2}\ ,
\end{equation}
which is the standard  magnon  energy on the spin chain side
or the leading term in the  BMN excitation energy  on the string  side.

Let us also  compute the  difference  of spins
\begin{equation}
J_1-J_2=\int_{0}^{2\pi}\frac{d\sigma}{2\pi}  \cos
2\psi\ \dot{\alpha}
\end{equation}
on the fluctuations  around the vacuum $\psi=0$, $\varphi=0$. Here
$\alpha=\mathcal{J}t$ is the ``fast''  coordinate, and $
\cos2\psi=\sqrt{1-n_{a}^2}=1-z_{a}^2$ so that
(to all orders in fluctuations)
\begin{equation}
J_{1}-J_{2}=
J-2 \sum_{n=-\infty}^{\infty}a_{n}^{\dagger}a_{n}\ ,
\end{equation}
where we again assumed  normal ordering.\foot{Here the normal ordering is
again equivalent to the   $\zeta$-function regularization with
$\sum_{n=-\infty}^{\infty} n^s=0$, \ $s=0,1,2,...$.}
Applied to $M$-impurity state
 %Introducing the number of impurities  operator
%\be
%\hat{M }=\sum_{n=-\infty}^{\infty}a_{n}^{\dagger}a_{n} \ee
 the above
relation gives $J_1-J_2=J-2M$. Since  $J_{1}+J_{2}=J$ we have
\begin{equation}
J_1=J-M\ ,\ \ \ \ \ \  \quad J_{2}=M \ .
\end{equation}
The corresponding gauge-theory
states are Tr$(\Phi_1 ^{J_{1}}\Phi_2 ^{J_{2}})+ ...$,
and $J$ plays the role  of the length of spin chain and $M$ is the number
of magnons.

%As introduced in the LL action, $J$ is indeed
%the length of the spin chain.

\subsection{$\tilde{\lambda}/J$ correction}

To compute  the $1/J$ correction  to the energy of $M$-impurity state
 one needs to include  the quartic term in the Hamiltonian \rf{quartic}
 or in
 $\bar H_4\equiv  \int_{0}^{2\pi}\frac{d\sigma}{2\pi} H_4$
 and use the standard  quantum mechanical perturbation theory.
Written in  terms of the creation and annihilation operators
 it has the form
\begin{eqnarray}\label{ham4}
\bar H_4 &=&\frac{\tilde{\lambda}}{4J}\sum_{n,m,k,l}n k
\bigg[\frac{1}{2}\delta_{n-k+m-l}(a_{n}a_{k}^{\dagger}a_{m}a_{l}^{\dagger}+
a_{n}^{\dagger}a_{k}a_{m}^{\dagger}a_{l})\nonumber\\
&-&\frac{1}{2}\delta_{n-m-k+l}(a_{n}a_{m}^{\dagger}
a_{k}^{\dagger}a_{l}+
a_{n}^{\dagger}a_{m}a_{k}a_{l}^{\dagger})\nonumber\\
&-&\delta_{n-m+k-l}(a_{n}a_{m}^{\dagger}a_{k}a_{l}^{\dagger}+
a_{n}^{\dagger}a_{m}a_{k}^{\dagger}a_{l})\bigg]\,.
\la{hea}\end{eqnarray}
Here we have  omitted  the time dependent
phases  ($a_n \to e^{- i \omega_n t} a_n $)  in the interacting Hamiltonian
$\bar H_{int}=\bar{H}_{4}+\bar{H}_{6}+...$  since they can be removed by
a unitary transformation with the quadratic Hamiltonian
$\bar{H}_{2}$. Here and below the summations over $n,m,...$ are
from $-\infty$ to $\infty$. One should decide about the
regularization, i.e. about the ordering of the operators $a_n$ and
$a^\dagger_n$. The natural choice is again the normal
ordering.\foot{It can be justified  by the requirement that both
the vacuum  state  $|0\rangle$ {\it and}  the 1-impurity state
$(a^\dagger_n +a^\dagger_{-n} ) |0\rangle$ should not receive subleading
corrections  to their energies.}
 The  part of the
 %AT
Hamiltonian  relevant  for  computing its  expectation value  in  a state
satisfying  the momentum constraint
is then
\begin{equation}
\bar H_4=-\frac{\tilde{\lambda}}{J} \sum_{n,m}nm
a_{n}^{\dagger}a_{m}^{\dagger}a_{n}a_{m} \ . \la{hi}
\end{equation}
Let us note that using the $\zeta$-function regularization
  we would instead obtain
from \rf{hea}
\begin{equation}
\bar H'_4=-\frac{\tilde{\lambda}}{J}\left(\sum_{n,m}nm
a_{n}^{\dagger}a_{m}^{\dagger}a_{n}a_{m}+ \sum_{n}n^2
a_{n}^{\dagger}a_{n}\right)\ .
\end{equation}
Thus at quartic oscillator  order the $\zeta$-function
regularization is not equivalent to normal ordering; we would still need
then to discard the $\sum_{n}n^2 a_{n}^{\dagger}a_{n}$ term that
would  shift the energy of the 1-impurity state. From full string
theory  computation perspective such term  should be cancelled by
the contribution of fermions (cf. \cite{parn,callan}).

Using \rf{hi}
 the correction to the energy of  $M$-impurity state is
found to be\foot{There is a degeneracy since there are states with
$\sum_{j=1}^{M}n_{j}=0$ that have
 the same energy
$\frac{\tilde{\lambda}}{2}\sum_{j=1}^{M}n_{j}^2$. The sets of
states $\{n_{i=1,...,M}\}$ and $\{n'_{i=1,...,M}\}$ satisfying this
condition are permutations of one another, but since in this paper we assume
 that $n_i$ are all different, these are the same
states. The only remaining degeneracy is a double degeneracy of
states $\{n\}$ and $\{-n\}$. One can see that the Hamiltonian
expectation value matrix is diagonal. This remains true also when
computing next order corrections.}
\begin{equation}
\langle M|\bar H_4 |M\rangle= \frac{\tilde{\lambda}}{J}\sum_{j=1}^{M}n_{j}^2\ .
\la{hw}
\end{equation}

\subsection{$\tilde{\lambda}/J^2$ correction}

Let us now consider the $O(1/J^2)$ correction to the energy
of $M$-impurity state.
Within the standard quantum-mechanical perturbation theory
for the continuum LL Hamiltonian
it should be given by the the sum of the  first order perturbation
term
 for $\bar H_6$ in \rf{six}  and the second order perturbation  term
  for $\bar H_4$ in \rf{hi}.
The latter  is
\begin{equation}\la{secc}
 \langle M| (\bar H_4)^{(2)} |M\rangle=
  \sum_{M\neq
M'}\frac{\langle M|\bar H_4 |M'\rangle \langle M'| \bar H_4 |M\rangle}{E_M-E_{M'}}
\end{equation}
where $|M'\rangle $ is any possible intermediate state, and
 $|M\rangle=a^\dagger _{n_{1}} ... a^\dagger_{n_{M}}|0\rangle$.

 Since $\bar H_4$ in \rf{hi} contains only terms of  the
form $(a^{\dagger})^2a^2$, the only non-trivial intermediate
states can be the $M'$-particle ones of the form
$ a^\dagger _{n'_{1}} ... a^\dagger_{n'_{M}}|0\rangle  $
with $M'=M$.
 Then in order for
the matrix element
$\langle M|\bar H_4 | a^\dagger _{n'_{1}} ... a^\dagger_{n'_{M}}|0\rangle $ to be
non-zero, there should be  a $j$ and $k$ such that   $n'_j=n_j+q$ and $n'_k= n_k-q$, with all
other $n_i'=n_i$, $i\ne j,k$.  In order for $|M\rangle$ to be distinct from $|M'\rangle$, we require that
$0\ne q\ne n_k-n_j$.    With these conditions, we then find that
\begin{equation}
\langle M|\bar H_4 |M'\rangle = -\frac{\tl}{J}\left[n_jn_k+(n_j+q)(n_k-q)\right]=
-\frac{\tl}{J}\left[2n_jn_k+q(n_k-n_j-q)\right]\,,
\end{equation}
if $n_k\ne n_j$ and where $n_j+q$ and  $n_k-q$ are not equal to one of the other $n_l$'s.
The energy difference is
\begin{equation}
E_M-E_{M'}=\frac\tl2\left[n_j^2+n_k^2-(n_j+q)^2-(n_k-q)^2\right]=\tl q(n_k-n_j-q)\,.
\end{equation}
 If $n_j+q=n_l$,
and so $|M'\rangle$ has two impurities with the same momenta, then the matrix element is
\begin{equation}
\langle M|\bar H_4 |M'\rangle = -\frac{\sqrt{2}\,\tl}{J}\left[n_jn_k+n_l(n_j+n_k-n_l)\right]\,,
\end{equation}
and the energy difference is
\begin{equation}
E_M-E_{M'}=\tl (n_l-n_j)(n_k-n_l)\,.
\end{equation}
 Hence, we find that
\begin{eqnarray}\label{H42sum}
&&\langle M| (\bar H_4)^{(2)} |M\rangle\\
&=&\frac{\tl}{J^2}\sum_{j<k}\left(\frac12\sum_{q=-\infty\atop0\ne q\ne n_k-n_j}^\infty
\frac{[2n_jn_k+q(n_k-n_j-q)]^2}{q(n_k-n_j-q)}+\sum_{l\ne j\atop l\ne k}\frac{\left[n_jn_k+n_l(n_j+n_k-n_l)\right]^2}{(n_l-n_j)(n_k-n_l)}\right)\nonumber\,
\end{eqnarray}
where the factor of $\frac12$ compensates for  a  double counting over
$M'$ and the sum over $l$ compensates for a missing contribution from $M'$ states with oscillators
with the same momenta.

The sum over $q$ in (\ref{H42sum}) is divergent and needs to be  regularized.  To this end, we can
write the sum over $q$  as
\bea
&&\sum_{q=-\infty\atop0\ne q\ne n_k-n_j}^\infty
\frac{(2n_jn_k+q(n_k-n_j-q))^2}{q(n_k-n_j-q)} \nonumber\\
&&=\sum_{q=-\infty\atop0\ne q\ne n_k-n_j}^\infty
\left[q(n_k-n_j-q)+4n_jn_k+\frac{4n_j^2n_k^2}{q(n_k-n_j-q)}\right]\,.
\eea
Using the  $\zeta$-function regularization, the first term  inside the square brackets gives zero after
summing over $q$, while the
second term gives
\begin{equation}\label{rel1}
\sum_{q=-\infty\atop0\ne q\ne n_k-n_j}^\infty4n_jn_k\ =\  4n_jn_k[2\zeta(0)-1]=-8n_jn_k\,.
\end{equation}
  The last term inside the brackets gives a finite
contribution\foot{While we have
assumed that $n_j\ne n_k$, it is clear that we can still
regularize the first term in \rf{H42sum} if $n_j=n_k$ (although
\rf{H42sum} will have additional symmetry factors).  In this case,
the expression in \rf{rel2} would be $-\frac{4\pi^2n_j^4}{3}$.}
\begin{equation}\label{rel2}
\sum_{q=-\infty\atop0\ne q\ne n_k-n_j}^\infty\frac{4n_j^2n_k^2}{q(n_k-n_j-q)}=\frac{4n_j^2n_k^2}{n_k-n_j}\sum_{q=-\infty\atop0\ne q\ne n_k-n_j}^\infty\left(\frac1q-\frac1{q-n_k+n_j}\right)=-\frac{8n_j^2n_k^2}{(n_k-n_j)^2}\,.
\end{equation}

The sum over $l$ in (\ref{H42sum}) can be symmetrized with the sum over $j$ and $k$, leading to the relation
\begin{eqnarray}\label{rel3}
&&\sum_{j<k}\sum_{l\ne j\atop l\ne k}\frac{\left[n_jn_k+n_l(n_j+n_k-n_l)\right]^2}{(n_l-n_j)(n_k-n_l)}
=
-\sum_{j<k<l}(n_j^2+n_k^2+n_l^2-n_jn_k-n_kn_l-n_ln_j)\nonumber\\
&&=-\frac{(M-1)(M-2)}{2}\sum_j n_j^2
+(M-2)\sum_{j<k}n_jn_k=-\frac{M(M-2)}{2}\sum_j n_j^2\,,
\end{eqnarray}
where we made use of the zero-momentum constraint in (\ref{momcond}).
Putting (\ref{rel1}), (\ref{rel2}) and (\ref{rel3}) into (\ref{H42sum}), we find that
\begin{equation}\label{H42}
\langle M| (\bar H_4)^{(2)} |M\rangle=-\frac{2\tl}{J^2}\sum^M_{j\ne
k}\frac{n_jn_k(n_j^2-n_jn_k+n_k^2)}{(n_j-n_k)^2}
-\frac{\tl M(M-2) }{2J^2}\sum^M_j n_j^2\,.
\end{equation}

Next, let us  consider  the expectation value of the six-order term
in the  LL Hamiltonian \rf{six}.
Expressing it  in terms of creation and annihilation operators using
\rf{lo},\rf{loo}   we face the question of regularization or how to
order the operators $a_n$ and $a^\dagger_n$.
It is no longer obvious that normal ordering is the right
prescription:  one could also keep the terms like $a^\dagger a^\dagger
a a $ and still  satisfy the requirement that the energy of
the vacuum and 1-impurity states is not shifted.
For example, if we use the $\zeta$-function prescription we get
\begin{equation}\la{pu}
\bar H_6
=-\frac{\tilde{\lambda}}{2J^2}\bigg[
\sum_{n,m,k}(nm-n^2)a_{n}^{\dagger}a_{m}^{\dagger}
a_{k}^{\dagger}a_{n}a_{m}a_{k}+ c \sum_{n,m}
2(nm-n^2)
a_{n}^{\dagger}a_{m}^{\dagger}a_{n}a_{m}\bigg]\ ,
\end{equation}
where  $c=1$.
On the other hand,  using normal ordering would give $c=0$, i.e.
the second $a^\dagger a^\dagger a a $ term  would be absent.
 It is unclear {\it a priori}  which  should be the right
value of $c$, i.e. if there are additional global
 conditions like protection of the energy of BPS states that should be imposed
  to match  either the
 gauge theory or the string theory results (which were not yet proven
 to be equal at this  order).
 We will see below that $c=0$ ({\it i.e.} normal ordering)
 reproduces  the correct spin-chain result, but for the moment
 let us  keep $c$ arbitrary.

For the  $M$-impurity state we have
\begin{equation}
\langle M|\sum_{n,m,k}nma_{n}^{\dagger}a_{m}^{\dagger}
a_{k}^{\dagger}a_{n}a_{m}a_{k} |M\rangle=-(M-2)\sum_{j=1}^{M}n_{j}^2\ ,
\end{equation}
\begin{equation}
\langle M|\sum_{n,m,k}n^2 a_{n}^{\dagger}a_{m}^{\dagger}
a_{k}^{\dagger}a_{n}a_{m}a_{k} |M\rangle=(M-1)(M-2)\sum_{j=1}^{M}n_{j}^2\ ,
\end{equation}
\begin{equation}
\langle M|\sum_{n,m}nma_{n}^{\dagger}a_{m}^{\dagger} a_{n}a_{m}
|M\rangle=-\sum_{j=1}^{M}n_{j}^2\ ,
\end{equation}
\begin{equation}
\langle M|\sum_{n,m}n^2a_{n}^{\dagger}a_{m}^{\dagger} a_{n}a_{m}
|M\rangle=(M-1)\sum_{j=1}^{M}n_{j}^2\ ,
\end{equation}
where we  used the momentum constraint in (\ref{momcond}).
Then we find  the following expectation value
\begin{equation}
\langle M| \bar H_6|M\rangle=\frac{\tilde{\lambda}}{2J^2}
  M(M- 2 + 2 c) \sum_{j=1}^{M}n_{j}^2\ . \la{ress}
\end{equation}

The sum of  \rf{H42} and \rf{ress} is  not yet the full result.
The original spin chain is discrete, so accordingly the coherent state (LL)
 action  in the continuum limit will include  also
higher derivative terms that are suppressed by powers of $J$.  In particular,
  at order $\l/J^4$  there is a  higher derivative  term
 in \rf{ii}
 \begin{equation}
\Delta L =  \frac{{  \tl \pi^2 }}{24J^2}
(\partial_{1}^{2} \vec n)^2 \ .
\end{equation}
It   should then  be  added
 as a quantum counterterm to  the LL action in
 order  to match the discrete spin chain result.
 Similar terms should  also appear   on the string-theory side,
 representing the effective  contributions of other bosonic and fermionic
 modes that are not included  directly in the LL   action, although details of how such
 terms can arise are   presently unclear.

Written in terms of the fields
$z_{a}$ or rescaled  fields  $f,g$ \rf{fg}
the leading  higher-derivative  correction to
the LL Hamiltonian is simply quadratic
\begin{equation}
\Delta {H}=-\Delta L=
 -\frac{\tilde{\lambda}\pi^2}{6J^2}(f''^2+g''^2) +
  O(\frac{\tilde{\lambda}}{J^4}) \ ,\ \ \ \ \ \
  \ \ \ \  \Delta \bar H_2 =
   -\frac{\tilde{\lambda}\pi^2}{6J^2}\sum_n  n^4 a^\dagger_n a_n
   \ ,
\end{equation}
 where we are again assuming normal ordering to avoid shifting the vacuum energy.
 Its expectation value  on the
$M$-impurity state is  found to be
\begin{equation}
\langle M|\Delta  \bar H_2  |M\rangle=-\frac{\tl \pi^2}{6
J^2}\sum_{j=1}^{M}n_{j}^4\ .  \la{eess}
\end{equation}

Combining the results of \rf{H42}, \rf{ress} and \rf{eess}, we find that the total $\tl/J^2$ correction is
\begin{equation}\label{E2Mfinal}
E^{(2)}_M=-\frac{\tl \pi^2}{6J^2}\sum_{j=1}^{M}n_{j}^4-\frac{2\tl}{J^2}\sum_{j\ne k}\frac{n_jn_k(n_j^2-n_jn_k+n_k^2)}{(n_j-n_k)^2}+\frac{\tilde{\lambda} cM}{J^2}
  \sum_{j=1}^{M}n_{j}^2
\end{equation}

In Appendix A we will  compute the $\tl/J^2$ correction to
 the spin-chain energy using directly  the Bethe ansatz;  we will  find that
eq. \rf{E2Mfinal} agrees with (\ref{E2final}) if $c=0$.  This means that the
normal ordering of the six-order  interaction
term $\bar H_6$ is the correct regularization
for reproducing  spin-chain results.

\subsection{$\tilde{\lambda}^2/J$ correction}

Let us now  include the next order $\tl^2$  term in the
classical LL action \rf{2loop} and  compute the leading $1/J$
correction to the energy of oscillator states.
 Putting again the LL  Lagrangian in the form (\ref{vv})
and rescaling the fields as in \rf{fg}
 we obtain the following fluctuation
Lagrangian to quartic order
\begin{eqnarray}
L=2\dot{f}g&-&\ha {\tilde{\lambda}}\bigg( f'^2+g'^2+
\frac{1}{J}\bigg[ 2 (g f'-f g')^2 -
 (f^2+g^2)
(f'^2+g'^2)\bigg]  \bigg) \nonumber\\
&+& { 1 \ov 8}
 {\tilde{\lambda}^2}\bigg( f''^2+g''^2+\frac{1}{J}\bigg[
(f'^2+g'^2)^2
+ 4(f f''+g g'')^2  \nonumber\\    &+&8(f f''+g  g'')(f'^2+g'^2)
- (f^2+g^2)(f f''''+g g'''')\bigg]
\bigg) \ .
\end{eqnarray}
The quadratic part of the Hamiltonian is
\begin{equation}
\bar H_2 =\frac{\tilde{\lambda}}{2}\int_{0}^{2\pi}\frac{d\sigma}{2\pi}(f'^2+g'^2)-\frac{\tilde{\lambda}^2
}{8} \int_{0}^{2\pi}\frac{d\sigma}{2\pi}(f''^2+g''^2)\ ,
\end{equation}
and the corresponding equations of motion are
\begin{equation}
\dot{g}=\frac{\tilde{\lambda}}{2}f''+\frac{\tilde{\lambda}^2}{8}f''''\ ,\  \ \ \ \ \ \
\quad
\dot{f}=-\frac{\tilde{\lambda}}{2}g''-\frac{\tilde{\lambda}^2}{8}g''''\
.
\end{equation}
Their solution is the same as in \rf{lo},\rf{loo},  now   with
\be
\omega_{n}= \ha \tl n^2 - {1\ov 8} \tl^2 n^4 \ , \ee
which is of course the expansion of the BMN frequency
$\omega_{n}=\sqrt{ 1 + \tl n^2} -1$ (see also \ci{krt}).
The normal-ordered quadratic Hamiltonian  is
then
\begin{equation}
\bar H_2 =    \sum_{n=-\infty}^{\infty}  \omega_{n}
a_{n}^{\dagger} a_{n} \ ,
\end{equation}
i.e. the leading term in the
 energy of the $M$-impurity state is
\begin{equation}
\langle M|\bar H_2 |M\rangle=\sum_{j=1}^{M} \omega_{n_{j}} =
\frac{\tilde{\lambda}}{2}\sum_{j=1}^{M}n_{j}^2-\frac{\tilde{\lambda}^2}{8}
\sum_{j=1}^{M}n_{j}^4\ .
\end{equation}
To find the $1/J$ correction we need to consider
the quartic term
$$\bar H_4=(\bar H_4)_{(1)}+ (\bar H_4)_{(2)}\ , $$
where  the first term is the
contribution of the
order $\tl$ term \rf{quartic} which was  already discussed
above in \rf{hw} while  the second term is the
 order $\tl^2$  contribution
\begin{eqnarray}
(\bar H_4)_{(2)}&=&-\frac{\tilde{\lambda}^2
}{8J}\int_{0}^{2\pi}\frac{d\sigma}{2\pi} \bigg[
(f'^2+g'^2)^2+4(f f''+g g'')^2\nonumber\\
&+&8(f f''+g g'')(f'^2+g'^2)-     (f^2+g^2)(f f''''+g
g'''')\bigg] \ .
\end{eqnarray}
Using  the normal-ordering prescription which
at the quartic order is again equivalent to the
condition that the energy of the vacuum and 1-impurity states is not
shifted  we finish with
\begin{equation}
 (\bar H_4)_{(2)}=\frac{\tilde{\lambda}^2}{J}
 \sum_{n,m}
n^3 m
a_{n}^{\dagger}a_{m}^{\dagger}a_{n}a_{m} \ .
%+\sum_{n=-\infty}^{\infty}
%n^{4}a_{n}^{\dagger}a_{n}\bigg]
\end{equation}
As a result,  the
subleading $\frac{\tilde{\lambda}^2}{J}$ contribution to the energy of an
$M$-impurity state is
\begin{equation}
\langle M|(\bar H_4)_{(2)}|M\rangle=
-\frac{\tilde{\lambda}^2}{J}\sum_{j=1}^{M}n_{j}^4\ . \la{iei}
\end{equation}

%%%%%%%%%%%%%%%%%%%%%%%%%%%%%%%%%%%%%%%%%%%%%%%%%%%%%%%%%%%%%%%%%%
\subsection{Comparison with gauge theory  and string theory results }
%%%%%%%%%%%%%%%%%%%%%%%%%%

Collecting the results \rf{hw},\rf{E2Mfinal},\rf{iei}
from the previous subsections
we find the following
expression for the  energy of the $M$-impurity state
$\{n_j\}$ satisfying (\ref{momcond}),
%as predicted by the quantum LL  approach
\begin{eqnarray}
E(M) = J &+&  \ha \tilde{\lambda}  \bigg[
\Bigg(1+\frac{2}{J}+\frac{2cM}{J^2}\bigg) \sum_{j=1}^M
n_{j}^2\nonumber\\
 &&\qquad - \frac{1}{J^2}\left(4\sum_{j\ne k}\frac{n_jn_k(n_j^2-n_jn_k+n_k^2)}{(n_j-n_k)^2}+\frac{\pi^2}{3} \sum_{j=1}^M  n_{j}^4\right)  + O ({1 \ov
J^3}) \Bigg]
\nonumber\\
&-&\ha \tl^2   \bigg[
\left (\frac{1}{4}+\frac{2}{J}\right)  \sum_{j=1}^M n_{j}^4     +O({1\ov J^2})\bigg] +
O ( \tl^3)  \,.
 \label{LLresult1}
\end{eqnarray}
Let us  compare this expression  with the gauge theory (exact spin chain)
 results. Here $J$ corresponds
 to  the number of spin chain  sites, i.e.  the spin chain length.
 Let us first consider the case of two
 impurities  ($n_1=-n_2=n$). In  the $SU(2)$ sector of SYM  theory
   the two-loop expression  for the anomalous dimension
   for $M=2$ operators  is known  exactly
   for finite   value of $J$
   \cite{bks}\footnote{The finite $J$ version of
    $M$-impurity energy
    was
derived recently in  the $SU(1|1)$ sector \cite{s}.}
\begin{equation}
E_{gauge} (M=2)  =J+\frac{\lambda}{\pi^2}\sin^{2}
\frac{\pi  n }{J-1}-\frac{\lambda^2}{\pi^4}\sin^{4}\frac{\pi
n}{J-1}\ \left(\frac{1}{4}+\frac{\cos^{2}\frac{\pi
n}{J-1}}{J-1}\right) +  O(\l^3) \ .
\end{equation}
Expanding in  large $J$ one gets
\begin{eqnarray}\label{Ecompare}
E_{gauge}(M=2)  =J&+&\tilde{\lambda} \bigg[
\left(1+\frac{2}{J}+\frac{3}{J^2}+...\right)n^2 -
\frac{\pi^2}{3J^2}n^4  + ... \bigg]
\nonumber\\
&-&{\tilde \lambda}^2\bigg[
\left(\frac{1}{4}+\frac{2}{J}+...\right)n^4  + ... \bigg] +  O(\tl^3)   \ .
\end{eqnarray}
We have already seen that the results match for the one-loop terms if the   6-order
 term in the oscillator Hamiltonian is normal ordered, corresponding to $c=0$.
Comparing \rf{Ecompare}  with the LL  result (\ref{LLresult1}) for $M=2$
  we also see that the $\tl^2/J$ terms match.

The string theory results for $1/J$ corrections with $M=2$
are known to agree with gauge theory at orders $\tl$ and $\tl^2$
\ci{callan}  and thus they  are also reproduced  by our
LL computation.  The $1/J^2 $  (order  ${\tilde{\lambda}}/{J^2}$)
term has  not yet been computed  directly on the string-theory side.

For arbitrary  $M$, the $1/J$ correction to the leading one loop
(order $\tilde{\lambda}$)  energy
 in the $SU(2)$ sector was found  from the
Bethe ansatz in  \cite{mz}. This expression was extended to
higher  orders in $\tl$ in \ci{bds,afs}.
 To second order  in $\tl$  the spin chain  result is
 (for different mode  numbers $n_j$)
\begin{equation}
E_{gauge}(M)=J+\frac{1}{2}\tilde{\lambda} \sum_{j=1}^M
\left(1+\frac{2}{J}\right) n_{j}^2
-\frac{1}{2}{\tilde\lambda}^2 \sum_{j=1}^M
\left(\frac{1}{4}+\frac{2}{J}\right) n_{j}^4 +O(\tilde{\lambda}^3) \ .
\label{stringBA}
\end{equation}
An equivalent $M$-impurity result  was found on the
string theory  side in \cite{swanson}.
Comparing to \rf{LLresult1} we  conclude that the
equal order $\tl$ and $\tl^2$ gauge and string
expression for $1/J$  corrections
is   thus   reproduced   by the
quantum LL computation  for  any number of impurities
$M$.\foot{
The string theory result of  \cite{swanson}
was presented in the form
$$
E=J'+M+\frac{{\lambda'}}{2}
\left(1-\frac{2M-2}{J}+...\right)\sum_{j=1}^{M} n_{j}^2
-\frac{{\lambda'}^2}{4} \left(\frac{1}{2}
-\frac{2M-4}{J}+...\right) \sum_{j=1}^{M}  n_{j}^4 +...
$$
where ${\lambda'}=\lambda/J'^2$.
To compare to the above  LL result  one needs to replace
  $J=J'+M$  and  then this expression  becomes equivalent to
  \rf{LLresult1}
  to the leading $1/J$ order.}

\bigskip

The order $\tl/J^2$ corrections for the $M$-impurity states in
\rf{E2Mfinal}, obtained from the quantum LL Hamiltonian, and \rf{E2final},
obtained from the Bethe ansatz, is a new result. A pressing issue is
to see if, and if, how, string theory reproduces it. If one
restricts the Hamiltonian used in the full string theory
\cite{callan} for obtaining the $\tilde{\lambda}/J$ corrections,
to the $SU(2)$ sector, one basically obtains the same Hamiltonian
as (\ref{hea}). This explains the agreement found for
$\tilde{\lambda}/J$ corrections as obtained from the full string
theory and the quantum LL approach.
Generalizing  to the $\tilde{\lambda}/J^2$
corrections we would expect that many of the details would be
similar to those in quantum LL computation, although
$\zeta$-function regularization would not be necessary because of
explicit cancelations of infinities due to supersymmetry (for
preliminary work in this direction see \cite{Swanson:2004mk}).

Let us stress that  getting  \rf{stringBA}  from the  LL action
did not  involve any non-trivial ambiguity apart from a normal
ordering assumption.  For the quartic term in the interacting
Hamiltonian, the normal ordering is required in order to protect
the  BPS states.   It would be interesting to see if a
corresponding principle can be found for the six oscillator term.

Another interesting computation that can be done is to find the
$\tilde{\lambda}^2/J^2$ corrections. This can be done using both the
quantum LL Hamiltonian  and the  Bethe ansatz, and we again expect to obtain the
matching.

\renewcommand{\theequation}{4.\arabic{equation}}
 \setcounter{equation}{0}
\section{The interacting Hamiltonian directly from the spin chain}
%%%%%%%%%%%%%%%%%%%%%%%%%%%%%%%%%%%%%%%%%%%

In this section we consider the Hamiltonian for interacting
magnons obtained
 directly from the Heisenberg  spin chain Hamiltonian.
Hence, we will skip the intermediary step of finding a long wave-length Lagrangian before quantizing.
By doing this we will encounter some subtleties in computing $1/J$ corrections.

In terms of  the spins, the Hamiltonian \rf{ferr}  for the one-loop $SU(2)$ sector is
\begin{equation}
H=\frac{\lambda}{8\pi^2}\sum_{\ell=1}^J
\left(\frac12-2\vec S_\ell\cdot\vec S_{\ell+1}\right)
\end{equation}
We can now write the spin operators in terms of auxiliary oscillators \cite{Faddeev:1996iy},
 $a_\ell$ and their  conjugates  $a^\dagger_\ell$,
as
\begin{equation}\label{su2osc}
S^3_\ell=\frac12-a^\dagger_\ell a_\ell,\qquad S^+_\ell=a_\ell,\qquad S^-_\ell=a^\dagger_\ell(1-a^\dagger_\ell a_\ell),
\end{equation}
where we assume  the usual commutation relations $[a_\ell,a^\dagger_{\ell'}]=\delta_{\ell,\ell'}$.
In terms of the oscillators, the Hamiltonian becomes
\begin{equation}\label{hamchain}
H=\frac{\lambda}{8\pi^2}\sum_{\ell=1}^J
\left[(a^\dagger_{\ell+1}-a^\dagger_{\ell})(a_{\ell+1}-a_{\ell})+(a^\dagger_{\ell+1}-a^\dagger_{\ell})^2
a_{\ell+1}a_{\ell}  \right].
\end{equation}

The Hilbert space for the spin chain is not the full Fock space of the oscillators: clearly,
 each site can be at most singly occupied.  Also,   the quartic term  in
(\ref{hamchain}) is not Hermitian,  nor are $S^+$ and $S^-$ conjugate to each other
under the usual conjugation rules
for the oscillators.   In order to have $H$ be Hermitian under the usual rules, we can add terms to
$H$ that are nonzero only for
states that have multiply occupied sites, which are outside of our Hilbert space.

In particular, let us first choose $H$ to be
\begin{eqnarray}\label{hamchain2}
H&=&\frac{\lambda}{8\pi^2}\sum_{\ell=1}^J
\Big((a^\dagger_{\ell+1}-a^\dagger_{\ell})(a_{\ell+1}-a_{\ell})-2a^\dagger_{\ell+1}a^\dagger_{\ell}
a_{\ell+1}a_{\ell}+a^\dagger_{\ell+1}a^\dagger_{\ell+1}a_{\ell+1}a_{\ell}+a^\dagger_{\ell}a^\dagger_{\ell}a_{\ell+1}a_{\ell}\nonumber\\
&&\qquad\qquad\ \ +a^\dagger_{\ell+1}a^\dagger_{\ell}a_{\ell+1}a_{\ell+1}+a^\dagger_{\ell+1}a^\dagger_{\ell}a_{\ell}a_{\ell}-a^\dagger_{\ell+1}a^\dagger_{\ell+1}a_{\ell+1}a_{\ell+1}-a^\dagger_{\ell}a^\dagger_{\ell}a_{\ell}a_{\ell}\Big),
\end{eqnarray}
where the last four terms in (\ref{hamchain2}) will not affect empty or singly occupied sites.  We can
now transform this to momentum space by defining
\begin{equation}
a_n=\sum_{\ell=1}^J e^{2\pi i n\ell/J} a_\ell\,,
\end{equation}
and thus getting  the expression
\begin{eqnarray}\label{hamchain3}
H&=&H^{(2)}+H^{(4)}=\frac{\lambda}{2\pi^2}\sum_{n=-J/2}^{J/2}\sin^2(\frac{\pi n}{J})a^\dagger_n a_n\\
&&\ \ -\frac{\lambda}{\pi^2J}
\sum_{n_1,n_2,n_3=-J/2}^{J/2}\sin(\frac{\pi n_1}{J})\sin(\frac{\pi n_3}{J})\cos(\frac{\pi (n_1-n_3)}{J})
a^\dagger_{n_1}a^\dagger_{n_2}a_{n_3}a_{n_1+n_2-n_3}\,.\nonumber
\end{eqnarray}
This Hamiltonian is exact.  If we take the large $J$ limit then we find
\begin{equation}\label{hamchain4}
H=\frac{\tl}{2}\sum_{n=-\infty}^{\infty}n^2a^\dagger_n a_n -\frac{\tl}{J}
\sum_{n_1,n_2,n_3=-\infty }^{\infty}n_1n_3
a^\dagger_{n_1}a^\dagger_{n_2}a_{n_3}a_{n_1+n_2-n_3}\,.
\end{equation}
We see that the quadratic part in  (\ref{hamchain4}) is consistent with (\ref{ham2}), but the quartic
part is not quite the same as (\ref{ham4}).   Beyond this order, $H$ in (\ref{hamchain4}) is obviously different from the LL Hamiltonian since the latter will have interaction terms of
all orders, as is required by integrability.

The reason for these differences is that the oscillator states here are not quite the same as those in
the LL case of the  previous section.
Since the Hamiltonian  $H$ in (\ref{hamchain4})
 is automatically normal ordered, it  will give the
 correct energy for the one magnon state, $a^\dagger_n|0\rangle$, \footnote{We temporarily ignore the momenum constraint.}
up to first order in $1/J$.  However, the second and higher order  results
will have corrections that depend on the lattice size.  If we next consider the two magnon state,
$a^\dagger_n a^\dagger_{-n}|0\rangle$,  we find that  the quartic piece in (\ref{hamchain4}) gives
no $1/J$ correction, which is in  conflict  with (\ref{hi}) as well as the exact answer given in
\cite{beisert,mz}.  The problem is that the two magnon state as written
above is not in the Hilbert space.
It turns out that there is a small overlap with states that have sites that are double occupied.\footnote{
Note that one way to avoid  going outside the Hilbert space is to use two
fermionic creation and annihilation operators at each site ({\it c.f. } \cite{beis}).  However,  it is not clear
how to compare the resulting Hamiltonian   to the LL Hamiltonian.}
Instead, we should choose for the two magnon state
\begin{equation}\label{2magnon}
|n,-n\rangle\equiv \frac1{\sqrt{1-{2\ov J}}}\bigg(a^\dagger_n a^\dagger_{-n}
-\frac1J\sum_{n'=-\infty}^\infty a^\dagger_{n'} a^\dagger_{-n'}\bigg)|0\rangle\ ,
\end{equation}
which has no overlap with states having multiply occupied sites.  With this definition of the two
magnon state and using the exact $H$ in (\ref{hamchain3}), we get
\begin{equation}
%label{ }
\langle n,-n|H^{(2)}|n,-n\rangle=\frac{\lambda}{\pi^2}\left[\frac{1-4/J}{1-2/J}\sin^2\frac{\pi n}J +
\frac2{1-2/J}\ \frac1{J^2}\sum_{m=-J/2}^{J/2 }\sin^2\frac{\pi m}J\right]\,.
\end{equation}
Performing the sum over the $J$ modes we find
\begin{equation}\label{expH2}
\langle n,-n|H^{(2)}|n,-n\rangle=\frac{\lambda}{\pi^2}\left[\frac{1-4/J}{1-2/J}\sin^2\frac{\pi n}J +
\frac{1/J}{1-2/J}\right]\,,
\end{equation}
where we see that the last term is of order $J\tl$.  We could have also approximated
$\sin\frac{\pi m}J\approx\frac{\pi m}J$ and done $\zeta$-function regularization of the sum where we would have found that the last term was absent.    To find the contribution from $H^{(4)}$
in \rf{hamchain3}, we note that
\begin{equation}
\langle0|a_m a_{-m}H^{(4)}a^\dag_n a^\dag_{-n}|0\rangle=-\frac{4\lambda}{\pi^2 J}\sin^2\frac{\pi m}J\sin^2\frac{\pi n}J\,,
\end{equation}
so that
%:
\begin{eqnarray}\label{expH4}
\langle n,-n|H^{(4)}|n,-n\rangle&=&\frac{\lambda}{\pi^2(1-2/J)}\Bigg[-4\sin^4\frac{\pi n}J +4\sin^2\frac{\pi n}J
\sum_{m=-J/2}^{J/2}\sin^2\frac{\pi m}J\nonumber\\
&&\qquad\qquad\qquad \ -\ \frac4{J^3}\bigg(\sum_{m=-J/2}^{J/2}\sin^2\frac{\pi m}J\bigg)^2\Bigg]\,.
\end{eqnarray}
This then gives
\begin{equation}\label{expH42}
\langle n,-n|H^{(4)}|n,-n\rangle=\frac{\lambda}{\pi^2(1-2/J)}\left(-4\sin^4\frac{\pi n}J +2\sin^2\frac{\pi n}J
-\frac1J\right)\,.
\end{equation}
Notice that the $\zeta$-function regularization of (\ref{expH4}) would have removed the last two terms in
(\ref{expH42}).

Combining (\ref{expH2}) with (\ref{expH42}), we arrive at
\begin{equation}
\langle n,-n|H^{(4)}|n,-n\rangle=\frac{\lambda}{\pi^2(1-2/J)}\left(\sin^2\frac{\pi n}J-4\sin^4\frac{\pi n}J\right)
\end{equation}
which is the desired result up to order $\tl/J$.   If we have used the $\zeta$-function regularization, we
would have canceled out the unwanted terms of order $J\tl$, but we would have gotten the
wrong $\tl/J$ term.

One can repeat the same for  more than two impurities,  although the computation  gets
 rather tedious.
Instead,  we will do  the  analysis   in a different  way, which will also allow us to compare
with the results in section 3.  We note that to order $1/J$, the
state $|n,-n\rangle$ can  be written  also  as
\begin{equation}
|n,-n\rangle=Ua^\dag_n a^\dag_{-n}|0\rangle\,,
\end{equation}
where $U$ is defined as
\begin{equation}
U=1-\frac1{2J}\left(\sum_{p,q,r}a^\dag_p a^\dag_{q}a_ra_{p+q-r}-\sum_{p,q}a^\dag_p a^\dag_{q}a_pa_{q}\right)   \,.
\end{equation}
The operator $U$ combines a projector with another factor
that properly normalizes the state.  In fact,  this same operator can be used on all multi-impurity states
to order $1/J$.   We can then
do a similarity transformation on $H$ and define a new Hamiltonian
\begin{equation}
\tH=U^{\dagger} HU\,,   \ \ \ \ \ \ \ \ \ \   U=U^\dagger \ .
\end{equation}
Clearly,  $\tH$ has the same action on the ground state as $H$.  It is also evident that $\tH$ is not normal ordered.  If we now expand   $\tH$
 to quartic order while normal ordering the operators, we find
\begin{equation}
\tH^{(4)}=\frac\tl {2J}\sum_{p,q}(p^2+q^2)a^\dag_p a^\dag_q a_p a_q-\frac\tl{J}\sum_{p,q,r}pr\,a^\dag_p a^\dag_q a_r a_{p+q-r}+{\rm O}({1\ov J^2})\,,
\end{equation}
where in the intermediate steps, the sums are carried out assuming that the oscillator momenta
run from $-J/2$ to $J/2$.
If we now consider the expectation value of $\tH^{(4)} $ for a multi-impurity state we find
\begin{equation}
\langle M |\tH^{(4)}|M\rangle=\frac{\tl}{J}\sum^M_{i<j}(n_i^2+n_j^2)-\frac{\tl}{J}\sum^M_{i<j}(n_i+n_j)^2
=-\frac{2\tl}{J}\sum^M_{i<j}n_in_j\,.
\end{equation}
Using the momentum constraint  $\sum_i  n_i=0$, we find the same result as (\ref{hw}).

Still, $\tH$ is not the same as in (\ref{hea}), again suggesting that there is a nontrivial map between
the oscillators in section 3 and those used  here.  Moreover, $\tH$ has a nonlocal part, indicating that
the transformation itself is nonlocal.

In any case, while this method of constructing the Hamiltonian is clearly more cumbersome than constructing it directly from the LL action, there are no ambiguities in normal ordering or regularization.
Computing the next-order correction is,  in principle,  doable and would   provide a useful check on
the results in (\ref{ress}).

\renewcommand{\theequation}{5.\arabic{equation}}
 \setcounter{equation}{0}

%%%%%%%%%%%%%%%%%%%%%%%%%%%%%%%%%%%%%%%%%%%%%%%%%%%
\section{$1/J$ correction to energy  of near BPS states   in the
$\beta$-deformed theory}
%%%%%%%%%%%%%%%%%%%%%%%%%%%%%%%%%%%%%%%%%%%%%%
 Let us   now apply the  approach  of section 3 of quantizing the LL
 Hamiltonian to obtain a new result:
 the $1/J$ correction to the energy of  BMN-type  states
 in the $\b$-deformed  version  of the AdS/CFT
 which  relates an exactly marginal superconformal
   deformation of SYM theory
 to string theory in  the  $AdS_5 \times (S^5)_\b$  background
 constructed using  T-dualities and coordinate shifts  \ci{lm}
 (we shall consider the case of real deformation parameter  $\b$).
 The corresponding string theory and  the spin chain Hamiltonian
 were  discussed in detail in \ci{frt1}, where the $\b\not=0$  analog
 of the LL action \rf{con},\rf{coon}  was derived
 both from the spin chain   and from the string action.
 The analog of the $SU(2)$ sector here  contains
  the  operators built out
 of two  chiral scalars, Tr$(\P_1^{J_1} \P_2^{J_2}) + ...$
 and the BPS vacuum is again the  $(J,0)$ state.
 The spectrum  of the  corresponding BMN states  was discussed
 in \ci{niar,lm,frt1}.

\subsection{Quantum LL   Hamiltonian  approach}

 Here our starting point will be the ``$\b$-deformed''  LL action
 found  in \ci{frt1} which generalizes  \rf{LLu}
 \begin{equation}
L=\cos 2\psi\  \dot{\varphi}-\frac{\tilde{\lambda}}{2}\left [\psi'^2+
\sin^2 2\psi\  ( \varphi'+\frac{1}{2}\bb ) ^2\right] \ , \ \ \ \ \ \ \
\bb \equiv \beta J \ .
\end{equation}
Here we ignored $O(\tilde{\lambda}^2)$ terms  and assumed that
in addition to $\tl$   the parameter
$\bb$ is  also  fixed in the  large $J$ limit (see \ci{frt1} for
details).
 This Lagrangian   can be rewritten also in terms of the    ``cartesian'' fields
  $n_a$ in \rf{laag} as
\begin{equation}\la{thi}
L=L_{_{\b=0}} (n)
-\frac{1}{8} \tilde{\lambda}\bar{\beta}^2 n^2  +
\frac{1}{4}\tilde{\lambda}\bar{\beta}  \epsilon_{ab} n'_a n_b \ ,
\end{equation}
where $L_{_{\b=0}} $ is the undeformed Lagrangian in \rf{laag}--(\ref{vector}).

We would like to expand \rf{thi}  near the ground state
with $n_a=0$  and compute the leading $1/J$  correction to the
fluctuation  spectrum. We shall  then compare this to the
result that can be found directly from the Bethe ansatz  equations
given in \ci{frt1}. The  full string-theory computation  of this
$1/J$ correction appears to be  rather complicated, but it should be
fairly clear by now that   it should  be correctly captured (to
the given leading order) by the quantum LL theory, so the string
and  gauge theory $\tl/J$  corrections should also match in the
$\b$-deformed theory.

The  basis  for this confidence is that  the LL computation of the
$\tl/J$  correction involves only the  quadratic and quartic
oscillator terms in the LL  Hamiltonian  and in this case
the choice of normal ordering prescription  appears to be essentially the unique
regularization option, leaving no ambiguity. This is confirmed by matching to
the exact spin  chain results.

Changing the variables  $n_{a}\to  z_{a}\to (f,g)$
as in \rf{nz},\rf{fg} and expanding near the $f=g=0$ vacuum
 we obtain the fluctuation Lagrangian up  to the quartic order
\begin{equation}
L=L_{_{\b=0}} (f,g)   -\frac{\tilde{\lambda}\bar{\beta}^2}{2}(f^2+g^2)
+\tilde{\lambda}\bar{\beta}(g
f'-f g')
+\frac{\tilde{\lambda}\bar{\beta}^2}{2J}(f^2+g^2)^2
-\frac{\tilde{\lambda}\bar{\beta}}{J}(g
f'-f g')(f^2+g^2)\ ,
\end{equation}
where $L_{_{\b=0}} (f,g)  $ is the undeformed fluctuation Lagrangian
in (\ref{qua}).
The linearized equations of motion (cf. \rf{eqmotion})
\begin{equation}
\dot{g}=\frac{\tilde{\lambda}}{2}f''-\tilde{\lambda}\bar{\beta}
g'-\frac{\tilde{\lambda}}{2} \bar{\beta}^2 f\ , \ \ \ \quad
\dot{f}=-\frac{\tilde{\lambda}}{2}g''-\tilde{\lambda}\bar{\beta}
f'+\frac{\tilde{\lambda}}{2} \bar{\beta}^2 g\ ,
\end{equation}
are solved by \rf{lo},\rf{loo}  with \be \la{wew} \omega_n = \ha
\tl ( n  + \bb)^2 \  . \ee As discussed  in \ci{frt1}, the
momentum constraint is unchanged  from its $\b=0$  form \rf{hou}.

The quadratic Hamiltonian  may be written as
\begin{equation}
\bar H_2 =\ha \tilde{\lambda}\int_{0}^{2\pi}\frac{d\sigma}{2\pi}\bigg[
(f' -  \bb g)^2 + (g' + \bb f)^2 \bigg]\ ,
\end{equation}
and upon quantization it becomes
\begin{equation}
\bar H_2=\sum_{n=-\infty}^{\infty} \omega_{n}  a_{n}^{\dagger}a_{n} \ .
\end{equation}
As in the  $\b=0$ case  here we used normal ordering to preserve
the  BPS ground state.

Applying this  Hamiltonian to a physical  $M$-impurity state
  we obtain for its energy
\begin{equation}
E =\frac{\tilde{\lambda}}{2}\sum_{j=1}^{M}(n_{j}+\bar{\beta})^2=
\frac{\tilde{\lambda}}{2}\sum_{j=1}^{M}n_{j}^2
+\frac{\tilde{\lambda}}{2}\bar{\beta}^2
M \ ,  \la{spe}
\end{equation}
where we used that
  $\sum_{j=1}^{M}n_{j}=0$ as follows from the momentum constraint.
As a result, the $\b$-dependent  part of the energy is sensitive
only to the number of impurities  but not to their detailed distribution.
Note also that the zero-mode states  with $n_j=0$  get non-trivial shifts.
  The same remarks will apply to the $1/J$ correction to
  the  leading spectrum \rf{spe}.

  The  quartic Hamiltonian is
\begin{equation}
\bar H_4=(\bar H_4)_{_{\b=0}}-\frac{\tilde{\lambda}}{2J}
\int_{0}^{2\pi}\frac{d\sigma}{2\pi}\bigg[   \bar{\beta}^2 (f^2+g^2)^2-
2  \bar{\beta}(f^2+g^2)(f' g-g' f)\bigg]\ ,
\end{equation}
where $\bar (H_4)_{_{\b=0}}$ is given by the integral of \rf{quartic}.
Assuming normal ordering, we find that
$\bar H_4$ can be written as (cf. \rf{hi})
\begin{equation}
\bar H_4=-\frac{\tilde{\lambda}}{J}\sum_{n,m}(n + \bb)(m + \bb)
a_{n}^{\dagger}a_{m}^{\dagger}a_{n}a_{m}\ .
\end{equation}
As a result, the leading quantum correction to the energy of the
 $M$-impurity state is found to be
\begin{equation}
\langle M|\bar H_4|M\rangle=\frac{\tilde{\lambda}}{J}\sum_{j=1}^{M}n_{j}^2-
\frac{\tilde{\lambda}\bar{\beta}^2}{J}(M^2-M)\ .
\end{equation}
The total  energy can be written  then as ($\sum_{j=1}^{M}n_{j}=0$)
\begin{equation}
E=J+\frac{\tilde{\lambda}}{2}\sum_{j=1}^{M}(n_{j}+\bar{\beta})^2\left(1+\frac{2}{J}\right)
-\frac{\tilde{\lambda}\bar{\beta}^2}{J}M^2
+ O( {\tl \ov J^2}, \tl^2) \ .
\label{betaLL}
\end{equation}

\subsection{Bethe ansatz approach }

Let us now show that the same expression can be found directly
from the Bethe ansatz  equations \ci{frt1} for the corresponding
anisotropic XXZ spin chain describing the 1-loop dilatation
operator \ci{roib} of the $\b$-deformed gauge theory in
 the $SU(2)_\b$ sector.
 The Bethe ansatz equations  given in \ci{roib,frt1}
 are
\be e^{-2\pi i\beta
J}\left(\frac{u_k+i/2}{u_k-i/2}\right)^J = \prod_{j\ne
k=1}^M \frac{u_k-u_j+i}{u_k-u_j-i} \ , \ee
\be
%\label{tw_Bethe}
%while the cyclicity conditions become:
%\begin{equation}
e^{-2\pi i\beta M}\prod_{k=1}^M\frac{u_k+i/2}{u_k-i/2}\=1\,,\ \ \ \ \ \ \
\
E_1= \frac{\lambda}{8\pi^2}\sum_{j=1}^M
\,{1\ov u_j^2 + 1/4}\,,  \ee
where $u_j$ are magnon rapidities and $J$ corresponds to the length of the chain.
To take the large $J$ limit one rescales the
 the
rapidities
$$
u_{k}=J x_{k} $$
and expands the  logarithm of the Bethe
equations in powers of $1/J$. This gives   to the
 leading order in $1/J$
\ci{frt1} \be \frac{1}{x_k} = \frac{2}{J}\sum_{\stackrel{j=1}{j\ne
k}}^M \frac{1}{x_k-x_j}+2\pi(n_k+\bar{\beta})\ ,\la{jo}  \ee \be
%\begin{eqnarray}
\sum_{k=1}^M  \frac{1}{x_k}=2\pi (mJ+\bar{\beta} M)~\ , \ \ \ \ \ \
E_1=\frac{\tilde \lambda}{8\pi^2}\,\sum_{j=1}^M {1\ov x^2_k} \,, \ \ \
\tl = {\l\ov J^2} \ ,
 \label{ener}
\ee
where we introduced again $\bar{\beta}=\beta J$.
In the BMN limit we are interested  in
$J$ is  taken to infinity  while $M$ held is fixed,
so that the integer $m$ should be set equal to zero.
Then we find  that
the perturbative solution of \rf{jo}
 for $x_k$  generalizing the leading-order one in
\ci{frt1}  is
\begin{equation}
x_{k}=\frac{1}{2\pi
(n_{k}+\bar{\beta})}\bigg(1-\frac{2}{J}\sum_{j\neq
k}^{M}\frac{n_{j}+\bar{\beta}}{n_{j}-n_{k}}\bigg) + O({ 1\ov J^2}) \ ,
\end{equation}
with   \rf{ener} implying that
\begin{equation}\la{oki}
\sum_{j=1}^{M}n_{j}=0 \ .
\end{equation}
As a result, the leading order
correction to the  dimension of the corresponding
$M$-impurity  BMN operators
is found to be\foot{Note that this expression cannot be obtained
from the $\b=0$ one by the shift  $n_k \to n_k + \bb$:
while the solution for $x_k$
can be generated this way,
the momentum   constraint must remain the same:  $\sum_j  n_j=0$.}
\begin{equation}
E_1 =\frac{\tilde{\lambda}}{2}\sum_{k=1}^{M}(n_{k}+\bar{\beta})^2
\bigg(1+\frac{4}{J}\sum_{j\neq
k}^{N}\frac{n_{j}+\bar{\beta}}{n_{j}-n_{k}}\bigg) + O( {\tl \ov J^2}) \ .
\end{equation}
Using   \rf{oki}    we get
\bea
E_1 &=&
\frac{\tilde{\lambda}}{2}\sum_{k=1}^{M}
n_{k}^2\left(1+\frac{2}{J}\right)
 + \ha  \tilde{\lambda} \bb^2  M
  \bigg[  1  - { 2(M-1) \ov  J } \bigg]
  \nonumber \\
&=&  \frac{\tilde{\lambda}}{2}\sum_{k=1}^{M}(n_{k}+\bar{\beta})^2\left(1+\frac{2}{J}\right)-
\frac{\tilde{\lambda}\bar{\beta}^2}{J} M^2
 \ ,
\eea
which is exactly   the expression found from the
LL action  \rf{betaLL}.
This provides  a nice illustration of the power
of the  approach  based on the quantum LL action.
We expect that one can generalize the arguments for the
 $1/J^2$ corrections in sec. 3.3 and Appendix A
to the case of nonzero $\beta$.

\renewcommand{\theequation}{6.\arabic{equation}}
 \setcounter{equation}{0}

%%%%%%%%%%%%%%%%%%%%%%%%%%%%%%%%%%%%%%%%%%%%%%%%%%%%%%%
\section{$1/J$  corrections to the energy
of   circular rotating string solution}
%%%%%%%%%%%%%%%%%%%%%%%%%%%%%%%%%%%%%%%%%%%%%%%%%%%%%%%%

In the previous sections we  used the LL action  to compute  energies
of fluctuations near the (constant)  vacuum  solution
$\vec n = (0,0,1)$  corresponding to the BPS vacuum  Tr$\P_1^J$   in gauge
theory and  the   $S^5$  geodesic in string theory.

In this section we extend the analysis of   section 3
 to expansions near a non-BPS  state
represented by the  simplest static
 solitonic solution of the  LL equations of motion
\begin{equation}
\psi=\frac{\pi}{4}\ ,\ \ \ \  \quad \varphi=m\sigma\ , \ \ \  {\rm i.e.} \
\ \  \
\vec n = (\cos 2m \s, \sin 2m \s , 0) \ . \la{sool}
\end{equation}
This LL solution   corresponds to the leading (order $\tl$)
term \ci{krt} in the  circular rotating string solution of \ci{ft2,art},
i.e. $ \XX_1 = {1 \ov \sqrt 2} e^{i w \tau + i m \s}, \
 \XX_1 = {1 \ov \sqrt 2} e^{i w \tau - i m \s},$
 with $w = \J = {J\ov \sqrt{\l}} $   and
 $J_1=J_2= J/2$.  Here $m$ is an integer winding number which we shall
assume to be  positive.\footnote{Note that
the limit of $m=0$  brings us back to the ($SO(3)$ rotated)
 vacuum solution $\vec n=(1,0,0)$.}
  The    classical
energy is
\begin{equation}
E_{0}=J+\frac{\lambda m^2}{2J}+O(\lambda^2)
=  J \bigg[ 1 +  \ha \tl m^2 +  O(\tl^2)\bigg] \ , \la{gra}
\end{equation}
i.e. is the leading term in the $\tl$  expansion of the energy
$E= \sqrt{J^2 + \l m^2}$ of the full circular string solution.

Our aim will be to  compute the quantum $1/J$  corrections to the
energy of this  non-constant  ground state  of the LL model (which
in this section we shall denote as $|0_m\rangle $). While in the vacuum
case of  section 3  the correction  to the ground state
energy was absent and we concentrated on computing corrections to
energies of  near-by  fluctuation modes (i.e. BMN states)   here
the question about
 the  form of $1/J$  corrections to this non-BPS ground state energy
 is already a non-trivial one.

To compute the quantum $1/J^n$   corrections
to the ground state energy \rf{gra}
one should expand the LL action \rf{LLu}
near the solution \rf{sool} and quantize  the
Hamiltonian  for the  fluctuations.
 A convenient starting point is the
  LL action  in the parametrization   given
 in \rf{fluct4}.  Introducing the
 two fluctuation fields $(f,g)$ as
 \be
 \varphi =  m\sigma + {1 \ov \sqrt{J}} f   \ , \ \ \ \ \ \ \ \ \ \ \ \ \ \
 \xi =   -\ha \sin (2 (\psi-{\pi\ov 4}))
 = {1 \ov \sqrt{J}} g \ , \la{fla}
 \ee
 we  find the following expression  for the fluctuation
Lagrangian up to the quartic order (cf. \rf{qua}--\rf{six})
\bea
L&=&2g\dot{f}
-  \frac{\tilde{\lambda}}{2}(f'^{2}+g'^{2}-4m^2
g^2)+\frac{4m \tilde{\lambda}}{\sqrt{J}}\  f' g^2
+ \frac{2\tilde{\lambda}}{J} \ g^2 (  f'^2-
g'^2)   + ... \nonumber\\
&=& 2g\dot{f}- (H_2 + H_3 + H_4 + ...)
 \ . \la{haaa}
\eea

\subsection{Leading   $\tl$   correction }

To compute the leading  ``one-loop'' correction to the ground
state energy one should consider the  quadratic order in
fluctuations and sum over the corresponding characteristic
frequencies  \ci{btz}. The  linearized equations of motion are
\begin{equation}
\dot{f}=-\frac{\tilde{\lambda}}{2}g''-2\tilde{\lambda}m^2 g\ ,\ \ \ \ \
 \quad
\dot{g}=\frac{\tilde{\lambda}}{2}f''\ ,
\end{equation}
and they are solved again as in \rf{lo},\rf{loo} where now
the characteristic frequency is
\be \la{oma}
\omega_{n}=
\frac{\tilde{\lambda}}{2}n\sqrt{n^2-4m^2} \equiv  \
 \pm \frac{\tilde{\lambda}}{2}n^2 w_n \ , \ \ \ \
 w_n\equiv \sqrt{1 - {4 m^2 \ov n^2} } \
, \ \ \   \   n= 0, \pm 1, \pm 2, .... \ . \ee
We see that this solution has  unstable (imaginary frequency)
fluctuation modes
with $n=\pm 1, ..., \pm 2 m $ which is a manifestation of an  instability of the
full homogeneous circular string  solution \ci{ft2,art}.
In what follows we shall formally ignore this  instability issue
(see \ci{fpt,ptt,btz} for discussions)
by formally   defining all the expressions by analytic continuation from
the region $m < 1/2 $.
Most of what follows can be
readily
repeated for a  very similar stable  $(S,J)$ solution of \ci{art}
appearing in the $SL(2)$ sector (with the corresponding LL action
 derived
in \ci{st,castell}) for which the  full 1-loop  quantum  string
correction was found in \ci{ptt} (see also \ci{szz}). The
discussion of the  simpler  $SU(2)$ solution has  an advantage of being
more  explicit.

Imposing the commutation relations \rf{opi} as dictated by \rf{haaa}
 we find that in terms of  canonically  normalized
 creation and annihilation operators with respect to
 our non-trivial  ground state
 $|0_m\rangle $
 $$ [a_{n},a_{k}^{\dagger}]=\delta_{n-k} \ ,  $$
   we have (cf. \rf{lo},\rf{loo})
\begin{equation}
f(t,\sigma)=\frac{1}{2}\sum_{n=-\infty}^{\infty} \sqrt{ w_n}
(a_{n}e^{-i\omega_{n}t+in\sigma}+
a_{n}^{\dagger}e^{i\omega_{n}t-in\sigma}) \ , \la{ho}
\end{equation}
\begin{equation}
g(t,\sigma)=\frac{1}{2}\sum_{n=-\infty}^{\infty}
 { 1 \ov \sqrt{ w_n} }   (-ia_{n}e^{-i\omega_{n}t+in\sigma}+i
a_{n}^{\dagger}e^{i\omega_{n}t-in\sigma})\ .  \la{oh}
\end{equation}
The quadratic Hamiltonian
in \rf{haaa} becomes
\begin{equation}
\bar H_2=  \int_{0}^{2\pi}\frac{d\sigma}{2\pi} H_2=
 \frac{1}{2}\sum_{n=-\infty}^{\infty}|\omega_{n}|(a_{n}a_{n}^{\dagger}+a_{n}^{\dagger}a_{n})
=\sum_{n=-\infty}^{\infty}|\omega_{n}|a_{n}^{\dagger}a_{n}+\frac{1}{2}\sum_{n=-\infty}^{\infty}
|\omega_{n}|\ . \la{jh}
\end{equation}
The leading quantum  correction to the energy of our solitonic solution is
then given by
\begin{equation}
E_1 =\langle 0_m|\bar H_2|0_m\rangle =\frac{1}{2}\sum_{n=-\infty}^{\infty}
|\omega_{n}|=
\frac{\tilde{\lambda}}{2}\sum_{n=1}^{\infty}n\sqrt{n^2-4m^2}\ . \la{sui}
\end{equation}
Here  we should not   of course use the normal ordering
prescription  since it simply ignores  the  shift of the vacuum energy.
Instead, as common  at quadratic oscillator level,
it is natural to define the divergent sum in \rf{sui}
using the $\zeta$-function regularization; this  gives
the following finite result  \ci{btz}
\begin{equation}
E_1
=\frac{\tilde{\lambda}}{2}\bigg[  m^2 +
\sum_{n=1}^{\infty}(n\sqrt{n^2-4m^2}-n^2+2m^2)
  \bigg]  \ . \la{ada}
\end{equation}
This is exactly the same expression  as found from the full
string-theory 1-loop computation \ci{ft3,fpt,ptt} as well as  from
the Bethe ansatz  on the spin chain side formally
applied to the corresponding one-cut  Bethe root distribution  \ci{btz,bfrey}. From the string
theory  point of view, the additional terms  in \rf{ada},  as
compared to the unregularized expression in \rf{sui},  come from
contributions of other bosonic and fermionic modes which make the
total string result finite.

\subsection{Subleading $\tilde{\lambda}/J$ correction}

Subleading $\tilde{\lambda}/J$ corrections
to the circular
string energy  (2-loop correction on the  string side
and $1/J^2$ correction on the spin chain side)
were not computed before. This  computation is, however,
 straightforward  in the present approach based on
  standard  perturbation theory
 for the quantum LL Hamiltonian in  \rf{haaa}.
Obviously, $\langle 0_m|\hH_3|0_m\rangle =0,$
so to compute  the $\tilde{\lambda}/J$ correction  to the ground state
energy
we  need to  consider the second order
perturbation theory term
 for $\hH_3$ and first order term
 for $\bar H_4.$

Written in  terms of creation and annihilation operators $\hH_3$
takes the form\footnote{Here again we remove the time dependent
exponential factors by a unitary transformation.}
\begin{eqnarray}
\hH_3&=&\frac{im\tilde{\lambda}}{2\sqrt{J}}\sum_{n,k,l=-\infty}^{\infty}
\sqrt{\frac{w_{k}}{w_{n}w_{l}}}\ k\bigg(a_{n}a_{l}a_{k}
\delta_{n+l+k}-a_{n}a_{l}a_{k}^{\dagger}\delta_{n+l-k}\nonumber\\
&+&a_{n}^{\dagger}a_{l}^{\dagger}a_{k}\delta_{n+l-k}
-a_{n}^{\dagger}a_{l}^{\dagger}a_{k}^{\dagger}\delta_{n+l+k}-
a_{n}a_{l}^{\dagger}a_{k}\delta_{n-l+k}+a_{n}
a_{l}^{\dagger}a_{k}^{\dagger}\delta_{n-l-k}\nonumber\\
&-&a_{n}^{\dagger}a_{l}a_{k}\delta_{n-l-k}+a_{n}^{\dagger}a_{l}a_{k}^{\dagger}
\delta_{n-l+k}\bigg)
\ ,  \end{eqnarray}
where $w_n$ was defined in \rf{oma}.
The second order perturbative correction
$\hH_3$ receives contribution only from the $3$-particle
intermediate  states
$$|\textbf{1}_{n}\textbf{1}_{k}\textbf{1}_{l}> = a^\dagger_n a^\dagger_k
a^\dagger_l |0_m\rangle  \ . $$
The contribution vanishes if $n=k=l$.
 When the values of
$n,k,l$ are all different we obtain \bea
\W_{1}&\equiv&-\frac{1}{3!}\sum_{n\not=k\not=l}\frac{\langle 0_m|\hH_3|
\textbf{1}_{n}\textbf{1}_{k}\textbf{1}_{l}><\textbf{1}_{l}
\textbf{1}_{k}\textbf{1}_{n}|\hH_3|0_m\rangle }{|\omega_{n}|+
|\omega_{k}|+|\omega_{l}|}
\nonumber \\
&=&-\frac{m^2 \tilde{\lambda}}{3 J}\sum_{n\not=k\not=l}\delta
(n+k+l) \frac{\left(l\sqrt{  {  w_{l} \ov w_{n}w_{k}  }     }+
n\sqrt{\frac{w_{n}}{w_{k}w_{l}}} +k \sqrt{\frac{w_{k}}{w_{n}w_{l}}
}\right)^2}{n^2 w_{n}+k^2 w_{k}+l^2 w_{l}} \label{W1} \ ,
\end{eqnarray}
where the sums are  from $-\infty$ to $\infty$.
When two values among  $n,k,l$ are equal the contribution
is
\begin{eqnarray}
\W_{2}&\equiv&- \sum_{n\neq
k}\frac{<0_{m}|\hH_3|\textbf{1}_{n}\textbf{2}_{k}>
<\textbf{2}_{k}\textbf{1}_{n}|\hH_3|0_m\rangle }{|\omega_{n}|+2|\omega_{k}|}
\nonumber\\
&=&- \frac{\tilde{\lambda} m^2}{J}\sum_{n\neq
k}\delta(n+2k)\frac{\left(2k\sqrt{\frac{1}{w_{n}}}+
n\sqrt{\frac{w_{n}}{w_{k}^2}}\right)^2}{n^2w_{n}+2k^2w_{k}}
\label{W2}\ .
\end{eqnarray}
To compute  $\langle 0_m| \hH_4| 0_m\rangle $   where  $\hH_4$  is the
integrated quartic term in \rf{haaa} we need   to use that
\begin{eqnarray}
\int_{0}^{2\pi}\frac{d\sigma}{2\pi} g^2 f'^2=&+&\frac{1}{16
J^2}\sum_{n,k=-\infty}^{\infty}\bigg[nk(2a_{n}^{\dagger}a_{k}^{\dagger}a_{n}a_{k}+
2a_{n}a_{k}a_{n}^{\dagger}a_{k}^{\dagger}+a_{n}a_{k}^{\dagger}a_{k}a_{n}^{\dagger}\nonumber\\
&+&a_{n}a_{k}^{\dagger}a_{n}^{\dagger}a_{k}+a_{n}^{\dagger}
a_{k}a_{n}a_{k}^{\dagger}+
a_{n}^{\dagger}a_{k}a_{k}^{\dagger}a_{n})+  k^2 \frac{
w_{k}}{w_{n}}(a_{n}a_{n}^{\dagger}a_{k}a_{k}^{\dagger}\nonumber\\
&+&a_{n}a_{n}^{\dagger}a_{k}^{\dagger}
a_{k}+a_{n}^{\dagger}a_{n}a_{k}a_{k}^{\dagger}+a_{n}^{\dagger}a_{n}a_{k}^{\dagger}a_{k})
\bigg]\ ,
\end{eqnarray}
\begin{eqnarray}
\int_{0}^{2\pi}\frac{d\sigma}{2\pi} g^2 g'^2=&-&\frac{1}{16
J^2}\sum_{n,k=-\infty}^{\infty}\bigg[\frac{nk}{w_{n}w_{k}}(2a_{n}^{\dagger}a_{k}^{\dagger}a_{n}a_{k}+
2a_{n}a_{k}a_{n}^{\dagger}a_{k}^{\dagger}-a_{n}a_{k}^{\dagger}a_{n}^{\dagger}a_{k}\nonumber\\
&-&a_{n}a_{k}^{\dagger}a_{k}a_{n}^{\dagger}-a_{n}^{\dagger}a_{k}a_{k}^{\dagger}a_{n}-
a_{n}^{\dagger}a_{n}a_{k}a_{k}^{\dagger})-\frac{
k^2}{w_{n}w_{k}}(a_{n}a_{n}^{\dagger}a_{k}^{\dagger}a_{k}\nonumber\\
&+&a_{n}a_{n}^{\dagger}
a_{k}a_{k}^{\dagger}+a_{n}^{\dagger}a_{n}a_{k}^{\dagger}a_{k}+a_{n}^{\dagger}a_{n}a_{k}a_{k}^{\dagger})
\bigg]\ ,
\end{eqnarray}
so that
\begin{eqnarray}\label{W3}
\W_3 &\equiv&\langle 0_m|\bar H_4|0_m\rangle \\
&=&-\ \frac{\tilde{\lambda}}{8J}\bigg(
\sum_{n=-\infty}^{\infty}\frac{n^2}{w_{n}^2}
-\sum_{n,k=-\infty}^{\infty} \frac{k^2}{ w_n w_{k}} +
3\sum_{n=-\infty}^{\infty}n^2
+\sum_{n=-\infty}^{\infty}\frac{1}{w_{n}}
\sum_{k=-\infty}^{\infty}k^2 w_{k}\bigg) \nonumber
\end{eqnarray}
where again $w_n = \sqrt{ 1 - { 4 m^2\ov n^2}}$.

As a result,  the  $\tilde{\lambda}/J$ correction to the energy is
 the sum of \rf{W1},\rf{W2},\rf{W3}
\begin{equation}
E_{2} = \W_{1}+\W_{2}+\W_{3} \ .  \la{gh}
\end{equation}
The  divergent  sum  $\W_3$
 can be again defined
using the $\zeta$-function regularization, but in contrast
to quadratic  oscillator case here it is no longer clear if this is the
regularization that  actually
corresponds to the  UV finite microscopic theories we are interested in  --
the string theory or  the spin chain (which we still expect to agree at this
order).

The total  energy to this  order  is then given by the sum of
\rf{gra},\rf{ada} and \rf{gh}, i.e.
\begin{equation}
E=J\left[1+\ha m^2 \tl \left(1 +\frac{c_{1}}{J}
+\frac{c_{2}}{J^2}+  O({1 \ov J^3}) \right) + O(\tl^2) \right]\ .
\end{equation}
The coefficients $c_{1}$ and $c_{2}$  given by
regularized  sums may
be  evaluated  numerically.
Taking  $m=1$  (and ignoring  imaginary contributions of unstable modes)
we found  that
\begin{equation}
 c_{1}=-0.892\ , \ \ \   \quad c_{2}=11.04 \ . \label{c2}
\end{equation}
Some details of evaluation of the corresponding sums are discussed
in Appendix B.

%%%%%%%%%%%%%%%%%%%%%%%%%%%%%%%%%%%%%%%%%%%%%%%%%
\subsection{$\tilde{\lambda}^2$ correction}
%%%%%%%%%%%%%%%%%%%%%%%%%%%%%%%%%%%%%%%%%%%%%%%%%%%%%%%

Let us now repeat  the computation of the leading correction to
the classical energy  from section 6.1  by starting with the LL
action \rf{2loop}  containing  an order $\tl^2$ correction  which
represents the  2-loop  contribution on the  gauge theory side or the
subleading term in the  expansion of the string action in the
large $\J$ limit. The solution  \rf{sool} remains the solution of
the LL equations corrected by higher-derivative  $\tl^2$
terms, while the classical energy \rf{gra}  gets an additional term $
- { 1 \ov 8}  J \tl^2  m^4 $. Written in angular parametrization
the  Lagrangian (\ref{2loop})  has the form
\begin{eqnarray}
L&=&\cos
2\psi\ \dot{\varphi} -\frac{\tilde{\lambda}}{2}\big(\psi'^2+
 \sin^2
2\psi\ \varphi'^2 \big)+
\frac{\tilde{\lambda}^2}{8}\bigg[\psi'^4+\psi''^2\nonumber\\
&+&(9+7\cos 4\psi) \psi'^2 \varphi'^2+\frac{1}{2}(5+3\cos 4\psi)
 \sin^2
2\psi \  \varphi'^4 + \sin^2
2\psi \ \varphi''^2   \nonumber\\
&-&2 \sin 4\psi \  \varphi' ( \psi'' \varphi'  -2 \psi'
\varphi'') \bigg]\ . \la{uuu}
\end{eqnarray}
Expanding it to quadratic order in fluctuations,  we get
\begin{eqnarray}
L&=&2g\dot{f}-\frac{1}{2}\tilde{\lambda} (f'^{2}+g'^{2}-4m^2
g^2)\nonumber\\  &+&
\frac{1}{8}\tilde{\lambda}^2 \big(f''^2+g''^2
+ 6 m^2 f'^2-6 m^2 g'^2+8 m^4 g^2 \big) \ .
\end{eqnarray}
The corresponding equations of motion are
\begin{equation}
\dot{f}=-\frac{\tilde{\lambda}}{2}g''
- 2\tilde{\lambda}m^2g
-\frac{\tilde{\lambda}^2}{8}g''''-\frac{3\tilde{\lambda}^2 }{4}
m^2 g'' -\tilde{\lambda}^2m^4g\ ,
\end{equation}
\begin{equation}
\dot{g}=\frac{\tilde{\lambda}}{2}f''
+\frac{\tilde{\lambda}^2}{8}f''''
-\frac{3\tilde{\lambda}^2}{4}m^2f''  \ .
\end{equation}
Their solution is again  given by \rf{ho},\rf{oh} where now
the characteristic frequencies are found to be\foot{Comparing to
\rf{ho},\rf{oh}
we set again  $\omega_n \equiv  \ha \tl n^2 w_n$.}
\begin{eqnarray}
\omega_{n}&=&\pm
\frac{\tilde{\lambda}}{2}\sqrt{
 (n^2-4m^2+\frac{3\tilde{\lambda}}{2} n^2
m^2-\frac{\tilde{\lambda}}{4}n^4 -2\tilde{\lambda}m^4)
(n^2-\frac{3\tilde{\lambda}}{2}n^2m^2-
\frac{\tilde{\lambda}}{4}n^4 )}
\end{eqnarray}
or, expanded in $\tl$ to the order we consider,
\begin{eqnarray}
\omega_{n}=
\frac{\tilde{\lambda}}{2} n \sqrt{n^2-4m^2}
- \frac{\tilde{\lambda}^2}{8} n (n^2 + 2 m^2) \sqrt{n^2-4 m^2}
  + O(\tilde{\lambda}^3) \ ,
\end{eqnarray}
where again  we formally  assume that  $n^2 > 4 m^2 $.
These are the same  characteristic frequencies
(the part that belongs  to the $SU(2)$ sector)
as found from  the full string analysis of fluctuations in \ci{ft3,art,fpt}.

The   leading
quantum  correction to the classical energy  to order $\tl^3$
is then
\begin{eqnarray}
E_{1} &=&\langle 0_m|\bar H_2|0_m\rangle
=\frac{1}{2}\sum_{n=-\infty}^{\infty}|\omega_{n}|
\nonumber\\
&=& \frac{\tilde{\lambda}}{2}\sum_{n=1}^{\infty} n \sqrt{n^2-4m^2}
-
\frac{\tilde{\lambda}^2}{8}\sum_{n=1}^{\infty}  n
( n^2 + 2 m^2 ) \sqrt{n^2- 4  m^2}  \ .  \label{2loopstatic}
\end{eqnarray}
The  sum in the  $\tl^2$  term  here
is divergent and needs to be regularized.
A natural regularization choice is again, as in \ci{btz},
to  subtract  and add  the divergent part and use the
  $\zeta$-function regularization for the latter.
  This gives a generalization of \rf{ada}:
  \begin{eqnarray}
E_{1} &=&  \frac{\tilde{\lambda}}{2}
\bigg[ m^2 + \sum_{n=1}^{\infty} \bigg( n \sqrt{n^2-4m^2}- n^2 +
 2m^2\bigg)
  \bigg]\nonumber\\
 &-&
\frac{\tilde{\lambda}^2}{8}
\bigg[ 3 m^4 +
\sum_{n=1}^{\infty} \bigg(
 n (  n^2  +2 m^2)  \sqrt{n^2-4  m^2}
 - n^4 +  6 m^4 \bigg)   \bigg]
     \ .  \label{koj}  \end{eqnarray}
Let us now compare this
regularized  expression
with the   result from string theory \ci{fpt}.
An immediate problem
is that while the expansion of the full
 finite  string  1-loop
result in powers of $\tl$ produces the  convergent
 expression \rf{ada}  at order $\tl$, the
 coefficient of the $\tl^2$ term
  happens to be given by a
 divergent sum over $n$ (this  has to do with
peculiar properties of the string  1-loop expression as a function of
$\tl$).
% suggesting that there should exist a
%better defined expression for it  than  in terms of a
% formal series over $n$, see also \ci{ptt,szz}.}
The string result is given by
 the zero-mode term and the infinite sum  of  string-mode contributions; by formally
 expanding both  parts in
powers of $\tl$ we get:
\begin{equation}
E_{1\ string}=E_{zero}+E_{non-zero}\ , \ \ \ \ \ \ \
E_{zero}=\frac{\tilde{\lambda}}{2}m^2-\frac{5\tilde{\lambda}^2}{8}m^4
+ O(\tl^3) \ , \la{zze}
\end{equation}
\begin{eqnarray}
E_{non-zero}&=&\frac{\tilde{\lambda}}{2}
\sum_{n=1}^{\infty}\bigg(n\sqrt{n^2-4m^2} - n^2
 +  2m^2\bigg)\nonumber\\
&-&
\frac{\tilde{\lambda}^2}{8}
\sum_{n=1}^{\infty} \bigg(
 n (n^2 +2 m^2) \sqrt{n^2-4  m^2}  -   n^4 +
  10 m^4 \bigg)  +   O(\tl^3) \ .\la{zzz} \eea
Compared to  \rf{2loopstatic} here  we get   extra  polynomial terms
 which represent contributions
of other string modes ``external'' to
 the $SU(2)$ sector.\foot{In essence,
the
 reason why
 we are able to match the string and LL results
 using a particular regularization is that contributions
 of all ``external'' modes happen to be  simply polynomial in $n$.}
The second,  order $\tl^2$,
 sum over $n$  is divergent, which is an artifact  of the
 naive  expansion of  the  UV finite  sum  over $n$
in powers of  $\tl$.
% (which assumes that $n$ is fixed at large $1/\tl$
%and thus ignores a ``tail'' of the sum).
 A
procedure for  extracting the coefficient of the $\tl^2$ term
turns out to be  equivalent
% this procedure  based on splitting the sum into the region of smaller $n$ and $n $ of order of $1/\tl$ and using an %integral representation for the latter.}
 to the
  $\zeta$-function regularization  prescription for \rf{zzz}\foot{We are grateful to N. Beisert
for an important explanation  related to  this point. The
$\zeta$-function regularization procedure does not apply at higher
orders in $\tilde{\lambda}$ \cite{bt,sz}.}. We then
 finish with exactly the same expression \rf{koj} as
found in  the $\zeta$-function regularized  LL model.
As was shown in \cite{szz} for a  similar $SL(2)$
sector solution, using  the $\zeta$-function regularized
expression instead  of formally divergent $\tl^2$  term in the
1-loop string energy \ci{ptt} one matches  the result   that
one finds from the Bethe ansatz\foot{Ref.\ci{szz}  considered the
``string'' Bethe ansatz  equations \ci{afs}, but to given order
$\tl^2$ they are the same as the gauge theory  Bethe ansatz
equations.}
 at the $1/J$ order; the same  statement
formally applies  also in the $SU(2)$ sector.

\bigskip
\bigskip

To conclude, the above discussion
indicates  that at quadratic oscillator level
the $\zeta$-function
 regularization of the quantum LL model
 is the right regularization prescription
 needed to reproduce the gauge theory and string theory results.
 The  ability to  obtain  both  the string and  gauge theory results from the
 quantum  LL  Hamiltonian   gives an ``explanation'' of their matching.
 The question  of which regularization should be used
   at quartic and higher interaction
 order  still remains  nontrivial. It is not a priori clear if the
 regularizations dictated by the string theory and the gauge theory sides will
 be the same, but we expect that the subleading
 $\tl/J$ and $\tl^2/J$   corrections will continue to agree between the two
 sides.

\bigskip

\renewcommand{\theequation}{7.\arabic{equation}}
 \setcounter{equation}{0}

\section{$1/J$ correction to the energy of  folded string}
%%%%%%%%%%%%%%%%%%%%%%%%

The circular solution discussed in the previous section was homogeneous:
derivatives of the background
fields were constant, and as a result the coefficients
in the fluctuation Lagrangian were also constant, leading to simple
algebraic equations for the characteristic frequencies.
The quantization of more general static but inhomogeneous
solutions like the   folded rotating  $(J_1,J_2)$  string
\ci{ft4,bmsz,bfst}  presents a challenge
since here finding the spectrum of the fluctuation
operator appears to be technically complicated
(cf. \ci{ft1}).
Considering only the LL sector of fluctuations
leads to a simplification, and, in view of the above discussion,
 should be  enough (at least to leading
order) for computing the 1-loop correction to the energy of
such an  inhomogeneous solution.

\subsection{Classical solution and fluctuations near it}

Starting with the LL action \rf{LLu}, the LL
solution we are going to
consider here  is the order $\tl$ part of the full $(J_1,J_2)$
folded string background, i.e. \ci{kru} \foot{Recall that
 $\varphi=(\varphi_{1}-\varphi_{2})/2,$  where
 $\varphi_{1}=w_{1}t,\ \varphi_{2}=w_{2}t$, so that
 $w= (w_{2}-w_{1})/{2}>0$ (we shall
assume that $w_{2}>w_{1}$). The
solution  describes a string located at the center
of $AdS$, while it is folded  and stretched along a big
 circle in $S^{5}$,
and rotates about its center of mass with frequency $w_{2}$.
 Its  center of mass
also moves (with frequency $w_{1}$) along an
 orthogonal big circle of  $S^{5}$.}
\be\la{fold}
\vp = - w t\ , \ \ \ \ \ \ \
\psi= \psi(\s) \ ,  \ee
 where  $\psi$ is a solution of the 1-d sine-Gordon equation:
\begin{equation}
\psi'' + 2\bw\sin 2\psi=0 \ , \ \ \ \ \ \ \ \ \
 \bw \equiv { w\ov \tl}\ ,   \label{SG1}
\end{equation}
i.e.
\begin{equation}
\psi'^2=2\bw (  \cos 2\psi-\cos 2\psi_0) \ ,
\label{SG}
\end{equation}
where  $\psi$  changes  from $-\psi_{0}$ to $\psi_{0}.$ The
solution can be expressed in terms of the  Jacobi elliptic
functions as:
\begin{equation}
\sin \psi(\sigma)=\sqrt{q}\  \sn (C\sigma,q),\ \ \ \ \  \quad
\cos
\psi(\sigma)=\dn (C\sigma,q) \ ,  \label{folded}
\end{equation}
\be  q=\sin^2 \psi_{0}\ , \ \ \ \ \
\sqrt{\bw}=\frac{1}{\pi}\KK(q)\ , \ \ \ \ \
C=\frac{2}{\pi}\KK(q)\ , \ \ \ \ \
\KK(q)=\int_{0}^{\frac{\pi}{2}}\frac{dx}{\sqrt{1-q \sin^2
x}}\
. \ee
Here $\psi_0,\bw$ and $A$ are functions of the parameter
 $q$ which itself
is related to the  ratio of the two spins $J_1,J_2$
through the integral \ci{kru}:
\begin{equation}
\frac{J_{1}-J_{2}}{J}=\int_{0}^{2\pi}\frac{d\sigma}{2\pi}\cos
2\psi \ , \ \ \ \ \ \   \ \ \ \ \ \ \ \ \ \   J= J_1 + J_2  \ .
\end{equation}
The order $\tl$ term in the
 classical  energy  can be expressed as \ci{bfst}
 \be
 E= J\bigg[ 1 + \tl  F_1 ({ J_2 \ov J})  + O(\tl^2) \bigg]\ ,
 \la{claas}\ee
 \be
 F_1 ({ J_2 \ov J}) = { 2 \ov \pi^2} \KK(q) [ \EE(q) - (1-q) \KK(q) ] \
 , \ \ \ \ \ q = q({J_2\ov J})\ , \ \ \
 {\EE(q)\ov \KK(q) } = 1 - {J_2\ov J}
  \ ,\la{spii}  \ee
 where $\EE(q)= \int_{0}^{\frac{\pi}{2}}dx \sqrt{1-q \sin^2
x}$. Expanding in small $q$ (small $\psi_0$, i.e. small  string size) or small $J_2/J$
we get \ci{ft2,kru}
 \begin{equation}
\alpha \equiv  \frac{J_{2}}{J}=\frac{q}{2}+\frac{q^2}{16}+O(q^3)
\ , \ \ \ \ \ \
q=2\alpha-\frac{\alpha^{2}}{2}+O(\alpha^3) \ .  \label{alpha}
\end{equation}
\begin{equation}
E_{0}=J\bigg[1 +\ha {\td\lambda} \a
\bigg(1+\frac{\alpha}{2}+\frac{3
\alpha^{2}}{8}+O(\alpha^{3})\bigg) +O(\tl^2)\bigg]  \ . \la{enek}
\end{equation}
To find the leading  $1/J$  correction  to the classical
energy we  need as in \rf{sui} to sum up the characteristic
frequencies of fluctuations near the above solution. Starting from
\rf{LLu} or \rf{fluct4}, the quadratic fluctuation Lagrangian is found to be
\be \la{flucy} L= 2 g   \dot f   - \ha \tl \bigg[  f'^2 + g'^2 -
  V_1(\sigma) f^2 -  V_2(\sigma)  g^2 \bigg]
\ , \ee
and the corresponding
 equations for  fluctuations
are
\begin{equation}
\dot{f}=-\frac{1}{2}[g''+V_{2}(\sigma) g]\ , \quad \ \ \ \ \
\dot{g}=\frac{1}{2}[f''+V_{1}(\sigma)f]\ , \la{eqs} \ee where the
potentials  depend on the elliptic function $\cos\psi(\sigma)=\dn
(A\sigma,q)$ in \rf{folded}: \be V_{1}(\sigma)= 12 \bw \cos
2\psi-8 \bw \cos 2\psi_{0}   \ ,\ \  \ \ \ \ \ \ V_{2}(\sigma)=
4\bw \cos 2\psi\ .  \la{pott}
\end{equation}
For notational simplicity
we  have rescaled the time by a factor of $\tl$;
this factor is easy to restore in the expressions for the characteristic
frequencies.

Since the potentials do not depend on time, to find the
characteristic frequencies \be  g\sim A_{1}(\sigma)e^{i\omega t}+
c.c. \ ,   \ \ \ \ \ \ \ \ \ \ \ \ f\sim  A_{2}(\sigma)e^{i\omega
t} + c.c.\ ,    \la{fgh}  \ee one is to solve the corresponding $
2 \times 2$ matrix Schrodinger equation on a circle, i.e. to find
the spectrum of the operator
\begin{eqnarray}\la{Q}
Q  = \left(%
\begin{array}{cc}
  0 &   -{ d^2 \ov d \s^2 } - V_{2}(\sigma)     \\
   { d^2 \ov d \s^2 } + V_{1}(\sigma) & 0    \\
\end{array}%
\right) \ .
\end{eqnarray}
The integrability of the  LL model
suggests  that this problem should have a systematic solution.\foot{The
general classical finite gap solution of the LL model is known
in terms  of $\theta$-functions
\ci{fadd-tah,kmmz}.
Linearizing it near the folded string solution one should be able
to extract, in principle,  the spectrum  of the operator $Q$.}
Being unable at present
to find the spectrum of $Q$  exactly,
we shall resort to perturbation theory
in string length $\psi_0$ or $q$,
or equivalently, in the ``filling fraction''
$ \alpha \equiv  \frac{J_{2}}{J}$.

%%%%%%%%%%%%%%%%%%%%%%%%%%%%%%%%%%%%%%%%%
\subsection{Short string expansion  }
%%%%%%%%%%%%%%%%%%%%%%%%%%%%%%%%%%%%%%%%

Let us start with
 the extreme short string limit   $ \psi_{0}\ll 1,$
 i.e.  $\a\to 0, \ q\to 0$.
Then $\sin \psi \approx \psi$ and the sine-Gordon equation
(\ref{SG1}) becomes linear
\begin{equation}
\psi''+ 4\bw \psi=0 \ , \ \ \
\end{equation}
with the solution
 \be
 \psi=\sqrt{q}\sin m \sigma\  [1 + O(q ) ]
 \ , \ \ \ \ \ \  \bw=\frac{1}{4}m^2  \ , \ \ee
 where $m$ is integer. The condition on $\bw$
 follows from the periodicity in $\s$.
 The integer $m$ represents the number of folds of the string.
Then the  potentials in \rf{pott}  are constant
\begin{equation}
V_{1}=V_{2}=m^2  + O(q) \ , \label{constpot}
\end{equation}
and we readily  find  the characteristic frequencies \be
\omega_{n}=\pm\frac{1}{2}\tilde{\lambda}|n^2-m^2| + O(q) \
.\la{freqq} \ee Note that up to an $n$-independent term these  are
the same as the BMN frequencies. Since  these frequencies depend
only on $\tl = {\l \ov J^2}$ and not on $J_2$, this limit
represents a nearly  point-like string and  the correction to the
ground-state  energy  should  vanish. This is  indeed what one
finds  using the $\zeta$-function regularization and observing
that since the Hamiltonian  in \rf{flucy} with $V_1,V_2$ in
\rf{constpot} is not positive definite,\foot{It is important to
stress that non-positivity of the quadratic fluctuation
Hamiltonian here does not imply an instability: the characteristic
frequencies are real. The folded string solution  is definitely
stable for small string length. The reason why we do not have an
instability as compared to, say, the inverted harmonic oscillator is
that here both the ``$p^2$'' and ``$q^2$'' terms
 in the
canonical Hamiltonian ($f$ and $g$ play the role of  momentum and
coordinate in \rf{flucy})
   may have opposite signs,
i.e.  one  may have a   ``ghost tachyon'', and  this just means
that the sign of the energy  changes, but energy remains real.}
 here the contributions  of the frequencies  of the
 $n^2 < m^2$ modes
 to the vacuum energy  enter with  a negative
 sign.
 This
 follows   also from the general prescription
 for the vacuum energy
   in terms  of characteristic frequencies
 of a mixed system of oscillators in \ci{blau} which was used in \ci{ptt}:
  \be \la{ei}
 E_{1} =  \fr{1}{2}\sum_{p=1}^{N} {\rm sign}(C_{p})\ \w_{p,0}
   +\fr{1}{2}\sum_{n=1}^\infty\;\sum_{I=1}^{2N} {\rm sign}
   (C_{I}^{(n)}) \ \w_{I,n}
 \ , \ee
 where
 \be \la{ccc}
C_{p}={2m_{11}(\w_{p,0})\w_{p,0} \prod_{q\neq
p}(\w_{p,0}^{2}-\w_{q,0}^{2})}\ ,\quad\ \ \ \ \ C_{I}^{(n)}=
{m_{11}(\w_{I,n})\prod _{J\neq I}(\w_{I,n}-\w_{J,n})} \ . \ee Here
$F^{T}( \om_{I,n},n)=F(- \om_{I,n},-n)$ is the matrix whose zero
determinant condition is used to find characteristic frequencies
and $m_{11}$ is the minor of $F$, i.e. the determinant of the
matrix obtained from $F$ by removing the first row and first
column. Since here the minor is essentially $m_{11}\sim(n^2-m^2)$
and its sign depends on $n$, it is important to use the above
general prescription with sign factors. As a result,  $E_1 = 0 +
O(q)$.

Let us now consider subleading  corrections. Setting  $m=1$  we
get from the small $q$
 expansion of the elliptic functions
in  the general solution  \rf{folded}
\begin{equation}
\sin \psi(\sigma)=\sqrt{q} \sin \sigma \left[1 +\frac{1}{4}
q\cos^2 \sigma+O(q^{2})\right] \ , \label{sol}
\end{equation}
\begin{equation}
\KK(q)=\frac{\pi}{2}\left(1+\frac{q}{4}+\frac{9q^2}{64}\right)+O(q^3)
\ , \ \ \ \ \ \ \ \bw=\frac{1}{4}\left(1+\frac{q}{2}+\frac{11
q^2}{32}\right)+O(q^3)\ .
\end{equation}
The potentials in \rf{pott}  become
\begin{equation}
V_{1}=1+\frac{9q}{2}-6q \sin^2 \sigma+O(q^{2}), \quad\ \ \ \
V_{2}=1+\frac{q}{2}-2q\sin^2 \sigma+O(q^{2})\ .  \label{pot}
\end{equation}
Then the eigenvalue problem for the fluctuation operator \rf{Q} or
the set of equations for the characteristic frequencies becomes
(see \rf{fgh}) \bea 2i\omega A_{1}&=&A_{2}''+(1+\frac{9q}{2}-6q
\sin^2 \sigma)A_{2}+ O(q^2) \ ,
\nonumber\\
-2i\omega A_{2}&=&A_{1}''+(1+\frac{q}{2}-2q \sin^2 \sigma)A_{1}+
O(q^2)  \ . \eea Combining these   we obtain the fourth-order
differential equation
\begin{equation}
  U A_1  = 4 \w^2 A_1
 \ , \ \ \ \ \ \ \ \
   U= U_0 + q U_1 + O(q^2) \ , \label{oper}
\end{equation}
where
   \be
U_0 = \frac{d^{4}}{d\sigma^{4}}+2\frac{d^{2}}{d\sigma^{2}}+1\ , \
\ \ \ \ \ U_1= (5-8\sin^2 \sigma)\frac{d^{2}}{d\sigma^{2}}-4 \sin
2\sigma\ \frac{d}{d\sigma}+ 1  \ .\la{pee}
 \ee
Since $A_{1}$ must be  periodic we can use  the Fourier expansion
$A_{1}=\sum_{n}c_{n}e^{in \sigma}, $ where  the coefficients then
satisfy
\begin{equation}
[n^4-n^2(2+q)+1+q- 4 \w^2 ]c_{n}-2q(n-2)(n-1)c_{n-2}- 2q
(n+2)(n+1)c_{n+2} =0 \ .  \label{sys}
\end{equation}
To leading order in $q$  this is simply
\begin{equation}
(n^4-2n^2+1 - 4 \w^2)c_{n}=0\ ,
\end{equation}
which gives of course the same leading-order expression as found
in \rf{freqq} for $m=1$, i.e. $\omega=\pm\frac{1}{2} (n^2-1)$
(here we ignore the factor of $\tl$ which was absorbed in $t$).

To compute the order $q$  correction to $\w$ we may use
perturbation theory. The unperturbed ($q=0$) operator is $U_0$  in
\rf{pee} and the
 unperturbed
 eigenvectors are
  $v_{n}=(...0,0,1,0,0,...)$ with $1$ at position $n$,
  with coordinate $\s$-space eigenfunctions
  $\langle \sigma|v_{n}>=e^{in\sigma}$.
   Then the  diagonal
matrix element of the perturbation operator $U_1$ in \rf{pee} in
this basis gives\foot{Each eigenvalue  of the unperturbed operator
(with the exception of  $n=0$ one) is double degenerate, i.e.  the
corresponding eigenfunctions are $e^{i\sigma n}, e^{-i\sigma n}.$
The $2\times 2$ matrix on these eigenvalues is diagonal.} $  <v_n
| U_1 |v_n > = - (n^2-1).$ As a result, we find
\begin{equation}
4\omega^2=(n^2-1)^2-q(n^2-1) \ , \ \ \
\end{equation}
so that to the   linear order in $q$
\begin{equation}
\omega_{n\not=\pm 1 }  =\pm \frac{1}{2}\left(n^2-1 -   \ha {q}
+O(q^2)\right) \ , \ \ \ \ \ \
 \  \w_{\pm 1} = 0 \ .   \label{freq1}
\end{equation}
Using these frequencies we find for
 the $\zeta$-function regularized leading  quantum correction
to the energy at order $O(q)$: \be E_{1}={ 1 \ov 4}
\tilde{\lambda}  \bigg[q  + O(q^2) \bigg] =   \ha \tl \bigg[\a  +
O(\a^2) \bigg]  \ ,   \ee where we used   \rf{alpha} and took into
account the sign factors in the general expression for the vacuum
energy in \rf{ei}.

Combining this quantum correction with the expression for the
classical energy \rf{enek} we get for the leading $1/J$ correction
to the energy of the  folded string:
\begin{eqnarray}\la{hii}
E= J &+&{\lambda \ov 2 J} \bigg[\frac{J_2}{J}
\left(1+\frac{1}{J}+O({1\ov J^2})\right)
\nonumber\\
&+& \ \ha ( \frac{J_2}{J})^2  \left(1+ \frac{a_2}{J} +
O(\frac{1}{J^2}) \right) +O( (\frac{J_2}{J})^{3})\bigg]  +
O({\l^2\ov J^3}) \ ,
\end{eqnarray}
or, equivalently,
\begin{equation}
E=J+\ha \tl  J_{2}\left(1+\frac{2+J_{2}}{2J} +O({J_2\ov J^2}
)\right)+O(\tl^{2}) \ .
\end{equation}
%For $J_{2}=2$, this matches the expression of the BMN energy in
%\cite{callan} for two impurities
Note  that this formally matches the expression for the near-BMN
correction in \rf{stringBA}  if we set $L=J, \ M=J_2=2$
\begin{equation}
E=J+  \tl \left(1+\frac{2}{J}+O({1\ov J^2})\right)+O(\tl^{2})\ .
\label{bmnfolded}
\end{equation}
This should  be expected since a short folded  string should be
 very close to a BMN state. This correspondence with the  BMN spectrum
 was known \ci{ft1}
 at the classical level where $J_2$  plays the role of the number of impurities $M$
 (assuming, e.g.,  that  all mode numbers $n_j$ are equal to 1); remarkably, it holds also at the quantum
 $1/J$ level.

 We  compute the $1/J$ correction
 at the next $( \frac{J_2}{J})^2$ order in Appendix C.
 Using again the $\zeta$-function regularization
  we  find the following  expression
 \be  a_2 = \frac{1}{2}-  {\pi^2 \ov 3} \approx - 2.79 \ . \ee
The discussion of
 the folded string solution in the $SU(2)$
sector can be repeated  for the folded string $(S,J)$ solution
\ci{ft1} in the $SL(2)$ sector; we give some details of this case
in Appendix D.

\bigskip

It would be interesting to compare these results for the leading
$1/J$ correction to the energy of spinning strings found from the
LL  model  with  the direct  Bethe ansatz and string theory
computations. We expect that as in the circular string case
discussed in \ci{btz},
 the two will match,
with the $\zeta$-function regularized  LL expression providing an
``explanation'' of their matching.

\bigskip

\section*{Acknowledgments }
%%%%%%%%%%%%%%%%%%%%%%%%%%%

We are  grateful to  N. Beisert, A. Mikhailov, S. Schafer-Nameki,
 R. Roiban, M. Zamaklar  and K. Zarembo
for useful discussions.  A.A.T.  is also grateful to  K.X. Vishik  for  helpful advice.
The work of J.A.M. was supported in part by the Swedish Research Council and by the
DOE under grant \#DE-FC02-94ER40818.
The work of A.T. and  A.A.T.
  was supported  by the DOE grant DE-FG02-91ER40690.  A.A.T.
acknowledges  the support of
 the INTAS grant  03-51-6346 and the RS Wolfson award.
 A.A.T.  also thanks KITP  for hospitality  while completing this work,
and acknowledges the partial  support of  NSF  grant  PHY99-07949
while  at KITP.

\renewcommand{\theequation}{A.\arabic{equation}}
 \setcounter{equation}{0}
\setcounter{section}{1} \setcounter{subsection}{0}
 \section*{Appendix A: $\tl/J^2$ corrections from the Bethe ansatz}

 In this appendix we compute the energies of an $M$-impurity state
 in the $SU(2)$ sector up to and including $1/J^2$ corrections.  Starting with the Bethe equations
 \begin{equation}\label{BES}
\left(\frac{u_i+i/2}{u_i-i/2}\right)^J=\prod_{j\ne i}^M\frac{u_i-u_j+i}{u_i-u_j-i}\,,
\end{equation}
 we can write (\ref{BES}) up to order $1/J^2$ accuracy as
\begin{equation}\label{BES2}
\exp\left[\left(\frac{i}{u_i}-\frac{i}{12u_i^3}\right)J\right]=\exp\left[\sum_{j\ne i}^M\frac{2i}{u_i-u_j}\right]\,.
\end{equation}
If we rewrite $1/u_i$ as
\begin{equation}
\frac1{u_i}=\frac{2\pi n_i}{J}+\Delta_i\,,
\end{equation}
then we find the equation
\begin{equation}
J\Delta_i-\frac{(2\pi n_i)^3}{J^2}=\frac{4\pi}{J}\sum_{j\ne i}^M\frac{n_in_j}{n_j-n_i}
+2\sum_{j\ne i}^M\frac{(\Delta_i n_j^2-\Delta_j n_i^2)}{(n_j-n_i)^2}\ .
\end{equation}
We will assume that all $n_i$ are different.
Up to the desired order, we can write\hfill\\
 \mbox{$\Delta_i=\Delta_i^{(1)}+\Delta_i^{(2)}$,} where
\begin{equation}
\Delta_i^{(1)}=\frac{4\pi}{J^2}\sum_{j\ne i}^M\frac{n_in_j}{n_j-n_i}\,,
\end{equation}
and
\begin{equation}
\Delta_i^{(2)}=\frac{(2\pi n_i)^3}{12J^3}+\frac{8\pi}{J^3}\left[\sum_{k\ne j\ne i\atop k\ne i}^M\frac{n_in_jn_k}{(n_j-n_i)(n_k-n_i)}+\sum_{j\ne i}^M\frac{n_in_j(n_i^2+n_j^2)}{(n_j-n_i)^3}\right]
\end{equation}

Now the energy is given by
\begin{eqnarray}
E&=&\frac{\lambda}{8\pi^2}\sum_i^M\frac1{u_i^2+1/4}\nonumber\\
&\approx&\frac{\lambda}{8\pi^2}\sum_i^M\left(\frac{2\pi n_i}{J}\right)^2\left[1+\frac{2\Delta_i J}{2\pi n_i}
+\frac{\Delta_i^2J^2}{(2\pi n_i)^2}\right]\left(1-\left(\frac{\pi n_i}{J}\right)^2\right)\,.
\end{eqnarray}
Hence as an expansion in $1/J$ we find
\begin{equation}
E^{(0)}=\frac\tl2\sum_i^M n_i^2\,,
\end{equation}
\begin{equation}
E^{(1)}=\frac{J\tl}{2\pi}\sum_i ^M\Delta^{(1)}_i n_i=-\frac{\tl}{J}\sum_{j\ne i}^Mn_in_j=\frac{\tl}{J}\sum_i^M n_i^2\,,
\end{equation}
where we used the momentum constraint \rf{momcond} in the last step, and
\begin{eqnarray}
E^{(2)}&=&-\frac{(2\pi)^2\tl}{8J^2}\sum_i^M n_i^4
+\frac{J\tl}{2\pi}\sum_i^M \Delta^{(2)}_i n_i+\frac{J^2\tl}{8\pi^2}\sum_i^M (\Delta^{(1)}_i)^2\nonumber\\
&=&-\frac{\pi^2\tl}{6J^2}\sum_i^M n_i^4+\frac{4\tl}{J^2}\left[\sum_{k\ne j\ne i\ne k}^M\frac{n_i^2n_jn_k}{(n_j-n_i)(n_k-n_i)}+\sum_{j\ne i}^M\frac{n_i^2n_j(n_i^2+n_j^2)}{(n_j-n_i)^3}\right]
\nonumber\\
&&\qquad\qquad\qquad\qquad
+\frac{2\tl}{J^2}\sum_{i}^M\sum_{j\ne i\atop k\ne i}^M\frac{n_i^2n_jn_k}{(n_k-n_i)(n_j-n_i)}\,.
\end{eqnarray}
Symmetrizing the sums, and splitting the last term into a piece where $k=j$ and another piece
where $k\ne j$, we find
\begin{eqnarray}\label{E2final}
E^{(2)}&=&-\frac{\pi^2\tl}{6J^2}\sum_i^M n_i^4
-\frac{2\tl}{J^2}\sum_{j\ne i}^M\frac{n_in_j(n_i^2-n_in_j+n_j^2)}{(n_j-n_i)^2}
\,.
\end{eqnarray}

\renewcommand{\theequation}{B.\arabic{equation}}
 \setcounter{equation}{0}
\setcounter{section}{1} \setcounter{subsection}{0}
 \section*{Appendix B: Evaluation of sums for  circular
  string}

Here we shall  provide
some details of the computation of the
coefficient $c_{2}$ in (\ref{c2}).
 We rewrite the sums
$\W_{1},\W_{2}$ in (\ref{W1}),(\ref{W2}) so that the summations are
from $0$ to $\infty$. To avoid
contributions from unstable modes we shall formally consider
only terms with $n > 2 m$  and  take the real part of the
series.  These complications are absent in the
similar  $SL(2)$ case  where the solution is stable,
but the $SU(2)$ case is useful for illustrating the
 computational procedure.
%(we shall set $m=1$ at the end).
Then
\begin{eqnarray}
\W_{1}&=&-\frac{2\tilde{\lambda}m^2}{3J}\bigg[\sum_{n\not=k>2m}^{\infty}a(n,k)+\sum_{n,k>2m}^{\infty}b(n,k)\bigg]
\ , \end{eqnarray}
with
\begin{equation}
a(n,k)=\frac{\left[k\sqrt{\frac{ w_{k}}{w_{n}w_{n+k}}}+
n\sqrt{\frac{w_{n}}{w_{k}w_{n+k}}}-(n+k)\sqrt{\frac{
w_{n+k}}{w_{n}w_{k}}}\right]^2}{n^2 w_{n}+k^2 w_{k}+ (n+k)^2
w_{n+k}}\ ,
\end{equation}
\begin{equation}
b(n,k)=\frac{\left[-k\sqrt{\frac{w_{k}}{w_{n}w_{n-k}}}+
n\sqrt{\frac{w_{n}}{w_{k}w_{n-k}}}-(n-k)\sqrt{\frac{
w_{n-k}}{w_{n}w_{k}}}\right]^2}{n^2 w_{n}+k^2 w_{k}+(n-k)^2
w_{n-k}}\ .
\end{equation}
Expanding  the above coefficients
 for large $n$ or large $k$ we find that they go as
$1/n^2$, and $1/k^2$, so  that the series are
convergent. Numerical computation gives,  for $m=1$,
\begin{equation}
\sum_{n\neq k>2}^{\infty}a(n,k)\simeq 0.33\ .
\end{equation}
To extract the real part  of the second series
$\sum b$   we split the sum into  two sums with $n> k+2m$
and $k> n+2m$. Then one can show that
\begin{equation}
\sum_{k,n>k+2m}^{\infty}b(n,k)+\sum_{n,k>n+2m}^{\infty}b(n,k)=2
\sum_{n,k>2m}^{\infty}a(n,k)\ .
\end{equation}
The sum  $\W_{2}$ in (\ref{W2})
can be rewritten as
\begin{eqnarray}
\W_{2}=-\frac{4m^2
\tilde{\lambda}}{J}\sum_{k>2m}^{\infty}\frac{\left(\frac{1}{\sqrt{w_{2k}}}-
\frac{\sqrt{w_{2k}}}{w_{k}}\right)^2}{w_{k}+2w_{2k}}=-\frac{2m^2
\tilde{\lambda}}{J}\sum_{n>2m}^{\infty}  a(n,n) \ .
\end{eqnarray}
Thus   the convergent sum
$\W_{1}+\W_{2}$  is
\begin{equation}
\W_{1}+\W_{2}=-\frac{2m\tilde{\lambda}}{J}
\bigg[\sum_{n,k>2m}^{\infty}a(n,k)+\frac{4}{3}\sum_{n>2m}a(n,n)
\bigg]
\ .
\end{equation}
Numerical evaluation for $m=1$ gives
\begin{equation}\la{ji}
\W_{1}=-0.76 \frac{\tilde{\lambda}}{J}\ , \ \ \ \ \ \ \ \ \
\W_{2}=-0.147 \frac{\tilde{\lambda}}{J}\ .
\end{equation}
The sum $\W_{3}$ in (\ref{W3})  is divergent;
 by using the $\zeta$-function regularization we get
\begin{equation}
\W_{3}=-\frac{\tilde{\lambda}}{4J}\left[S_{1}-5-10m^2-
2(S_{2}-\frac{5}
{2})
(S_{3}-S_{4}-10m^2)\right]\ ,
\end{equation}
where the convergent sums  $S_{k}$ are defined by
$$
S_{1}=\sum_{n>2m}^{\infty}\frac{16 m^4}{n^2-4m^2},\ \ \ \ \ \  \quad
S_{2}=\sum_{n>2m}^{\infty}\left(\frac{n}{\sqrt{n^2-4m^2}}-1\right)\
,
$$ $$
S_{3}=\sum_{n>2m}^{\infty}\left(\frac{n^3}{\sqrt{n^2-4m^2}}-n^2-2m^2\right),
\quad
S_{4}=\sum_{n>2m}^\infty\left(n\sqrt{n^2-4m^2}-n^2+2m^2\right).
$$
Numerical evaluation for  $m=1$
 gives
\begin{equation}
\W_{3}= 6.43 \frac{\tilde{\lambda}}{J}\ . \la{ki}
\end{equation}
Combining \rf{ji} and \rf{ki}
we get
\begin{equation}
E_{2}=\W_1+\W_2+\W_3=5.52  \frac{\tilde{\lambda}}{J}\ ,
\end{equation}
leading to the value of $c_2$ in \rf{c2}.

\renewcommand{\theequation}{C.\arabic{equation}}
 \setcounter{equation}{0}
  \section*{Appendix C: Subleading term
  in   folded string energy
   }
Here we extend the computation of quantum correction to the folded
string energy
 in section 7.2 to the next $q^2$
or $\a^2= ({ J_2 \ov J})^2 $ order. To this order the potentials
in \rf{pott}  are
\begin{equation}
V_{1}=1+\frac{9q}{2}-6q \sin^2 \sigma+\frac{75q^2}{32}-3q^2 \sin^2
\sigma\  (1+\cos^2 \sigma)+O(q^{3})\ ,
\end{equation}
\begin{equation}
V_{2}=1+\frac{q}{2}-2q\sin^2 \sigma+\frac{11q^2}{32}-q^2\sin^2
\sigma\ (1+\cos^2 \sigma)+O(q^{3}) \label{pot1}\ ,
\end{equation}
 and so the  $O(q^2)$ term  in \rf{oper} is   found to be
\begin{equation}
U_2 =\left(\frac{43}{16}-4\sin^2 \sigma\ (1+\cos^2 \sigma) \right)
\frac{d^{2}}{d\sigma^{2}} -2\sigma \sin 2\sigma \left(
 1 + \cos 2\sigma\right)\frac{d}{d\sigma}+\frac{15}{16}.
\end{equation}
We need  to use the  second-order perturbation theory in $q$  for
the operator \rf{oper}, i.e. to  combine  the first order term for
$U_2$ with second-order term for $U_1$. Let us denote the
eigenvalues as \be W_n= W_{n}^{(0)} + q W_{n}^{(1)} + q^2
(W'_{n}{}^{(2)} + W''_{n}{}^{(2)} ) +  O(q^3)  \ , \ee where,  as
found in section 6.2, $W_{n}^{(0)}= (n^2-1)^2, \  W_{n}^{(1)}=-
(n^2-1), $ and $W'_{n}{}^{(2)}$   comes from second-order term in
$U_1$  and $W''_{n}{}^{(2)}$ from $U_2$. Noting that
 the unperturbed
eigenvalues are double degenerate, we  find  that the second-order
perturbation theory corrections
 $ W'_{n}{}^{(2)}$ are
found  by solving the zero-determinant condition of the matrix
% \footnote{Note that the operator $O(q)=q(5-8\sin^2
%\sigma)\frac{d^{2}}{d^{2}\sigma}-4q \sin
%(2\sigma)\frac{d}{d\sigma}+q$ is not hermitian.}
\begin{eqnarray}
\left(%
\begin{array}{cc}
  \sum_{k\neq n, -n}\frac{<n|U_1|k><k|U_1|n>}
  {W_{n}^{(0)}-W_{k}^{(0)}}-W'_{n}{}^{(2)} &
  \sum_{k\neq n,-n}\frac{<n|U_1|k><k|U_1|-n>}
  {W_{n}^{(0)}-W_{k}^{(0)}} \\
  \sum_{k\neq n,-n}\frac{<-n|U_1|k><k|U_1|n>}
  {W_{n}^{(0)}-W_{k}^{(0)}} &   \sum_{k\neq n,
  -n}\frac{<-n|U_1|k><k|U_1|-n>}{W_{n}^{(0)}-
  W_{k}^{(0)}}-W'_{n}{}^{(2)}  \\
\end{array}%
\right)
\end{eqnarray}
This matrix is diagonal, so the degeneracy is not lifted also in
the second-order perturbation theory. Computing $W'_{n}{}^{(2)}$
we obtain:
\begin{equation}
W'_{n\neq 1 }{}^{(2)}=-\frac{n^2}{n^2-1}, \quad\ \ \ \ \
W'_{1}{}^{(2)}=-\frac{3}{4} \ .
\end{equation}
To find $W''_{n}{}^{(2)}$ we consider  the
 extension of (\ref{sys}) to the $q^2$ order
\begin{eqnarray}
&&0=[n^4-n^2(2+q)+1+q-4 \w^2]c_{n}-2q (n-2)(n-1)c_{n-2} \nonumber\\
&-&2q (n+2)(n+1)c_{n+2} - q^2 \bigg[
\frac{3n^2 - 15 }{16}c_{n} + (n-2)(n-1)c_{n-2}  \\
&+&  (n+2)(n+1)c_{n+2}+\frac{1}{4} (n-4)(n-2)c_{n-4} +\frac{1}{4}
(n+4)(n+2)c_{n+2}\bigg] + O(q^3). \nonumber
\end{eqnarray}
Proceeding as at leading order in $q$, we find that \be \la{hp}
W''_{n}{}^{(2)} = -\frac{ 3n^2-15}{16} \ .\ee Combining the above
expressions,
 we get that $\w=0$ for $n=\pm 1$ while
for $n\neq \pm1$
\begin{equation}
4\omega^2=(n^2-1)^2-q(n^2-1)- q^2 \bigg[ \frac{ n^2}{n^2-1}  +
\frac{1}{16}(3n^2 -15)\bigg]  + O(q^3) \ ,
\end{equation}
i.e. the characteristic frequencies are
\begin{eqnarray}
\omega_{n\not=\pm1}=\pm \frac{1}{2}\bigg[n^2-1 -\ha q
-\frac{1}{32} q^2 \frac{3n^4+2n^2+11}{(n^2-1)^2}+O(q^3)\bigg].
\end{eqnarray}
Computing their sum, we get for the quantum correction to the
energy (restoring the $\tl$ factor)
\begin{equation}
E_{1} =\tilde{\lambda}\left[\frac{q}{4}-\frac{11q^2}{128}-
\frac{q^2}{2}\sum_{n=2}^{\infty}\frac{3 n^4+2 n^2+11}{32
(n^2-1)^2}+O(q^3)\right] + O(\tl^2) \ .\la{kou}
\end{equation}
The sum here is divergent. Applying  again the $\zeta$-function
regularization ($\zeta(s)=\sum_{k=1}^{\infty}k^{-s}$), the sum can be easily computed\foot{We thank N. Beisert  for this
suggestion.}
  and using  the relation (\ref{alpha}), we end up with the
 following expression for the
1-loop quantum correction to order $O(\alpha^2)$
\begin{equation}
E_{1}=\ha \tl \left[{\alpha}+ \ha a_2
{\alpha^2}+O(\alpha^3)\right] \ ,\ee
\be   \ \ \ \ \ a_2 =
\frac{1}{2}-2\zeta(2)=  \ha -  { \pi^2 \ov 3}    \ ,
\end{equation}
which leads to the  expression in \rf{hii}.

\renewcommand{\theequation}{D.\arabic{equation}}
 \setcounter{equation}{0}
\setcounter{section}{1} \setcounter{subsection}{0}

  \section*{Appendix D: Folded string solution
  in  $SL(2)$ sector}

Here we shall repeat the discussion of section 7 for a folded
string solution  in the $SL(2)$ sector. The $SL(2)$ sector
describes strings rotating in $AdS_3$ part of $AdS_5$ and whose
center of mass is moving along big  circle of $S^5$, i.e. their
energy is  parametrized  by the two spins $(S,J)$. The fast string
limit corresponds to $J$ being large with $S/J$ and $\tl = {\l \ov
J^2}$  being fixed \ci{ft1}.

The corresponding  Landau-Lifshitz Lagrangian \ci{st} is similar
to the one in the $SU(2)$ case \rf{coon} with  $U_a \rightarrow
V_r $ ($ V_{r}^{*}V^{r}= -1$, $  V^r =\eta^{rs}V_s, \
 \eta_{rs}= (-1,1)$)
\begin{equation}\la{sl}
L=-iV_{r}^{*}\partial_{0}V^{r}-\frac{\tilde{\lambda}}{2}|D_{1}V_{r}|^2
+ O(\tl^2) \ .
\end{equation}
In  the parametrization
\begin{equation}
V_{0}=\cosh \rho\ e^{i\eta},\ \ \ \ \ \ \
 \quad V_{1}=\sinh \rho\ e^{-i\eta}\ ,
\end{equation}
where $\rho$ is the radial $AdS_5$ coordinate and $\eta=\ha
(t-\phi)$ the above LL Lagrangian \rf{sl} becomes
\begin{equation}
L=-\cosh 2\rho\ \dot{\eta}-\frac{\tilde{\lambda}}{2}
(\rho'^2+\sinh^2 2\rho\ \eta'^2 )\ .   \label{LLs}
\end{equation}
%\subsection{Folded string solution}
 The
folded string solution \cite{ft2}  describes a string which is
stretched in the radial direction $\rho$, rotates in a plane in
$AdS_5$  about its center of mass and also  moves  along a  big
circle in $S^{5}$. The string solution is given by $t=\kappa
\tau$, $\phi=\omega_{1}\tau$, $\vp_3= w_3 \tau$,
 and $\rho=\rho(\sigma)$.
 To leading order in the $1/\J$ expansion, the corresponding
 solution of the LL equations  is
$\eta=-w\tau$,  \  $w=\frac{\omega_{1}-\kappa}{2}$ (we shall
assume $\omega_{1}>\kappa$), and $\rho$ satisfies the equation:
\begin{equation}
\rho''  + 2\bw \sinh 2\rho=0\ , \ \ \ \ \ \ \ \ \ \ \
   \bw= {w\ov \tl} \ ,   \label{SG2}
\end{equation}
i.e.
\begin{equation}
\rho'^2=2\bw  ({\cosh}\ 2\rho_0 -{\cosh}\ 2\rho )\ ,
\end{equation}
with  $\rho$ changing  from $0$ to $\rho_{0}.$ As discussed  in
\cite{bfst,art},
 this folded  $SL(2)$ sector solution
is related to the folded solution (\ref{folded}) in the $SU(2)$
sector by the following analytic continuation
$$
\rho\rightarrow i\psi, \quad \eta\rightarrow \varphi,  \quad
\kappa \rightarrow w_{1}, \quad \omega_{1}\rightarrow w_{2}, \quad
w_{3}\rightarrow \kappa\ .
$$
Under this transformation the  equation (\ref{SG2}) becomes
(\ref{SG1}), and also the LL Lagrangian (\ref{LLs}) becomes
(\ref{LLu}) up to an overall  sign.

As in the $SU(2)$ case we may  consider quadratic  LL fluctuations
near the  folded string solution (we again rescale the time
coordinate by $\tl$)
\begin{equation}
\dot{f}=-\frac{1}{2}\left[g''+4\bw (3\cosh2\rho-2\cosh 2\rho_{0})\
g \right]\ , \ \ \ \ \dot{g}=\frac{1}{2}\left( f''+4\bw  \cosh
2\rho\ f \right)\ .
\end{equation}
These equations  are exactly the same as in the $SU(2)$ sector
\rf{eqs},\rf{pott} with $\rho\to i\psi$.

Here the short string limit  corresponds to $\rho_{0}\to  0.$ The
solution of (\ref{SG2}) in the small $\rho_{0}$ limit is (for
1-fold case  $m=1$):
\begin{equation}
\sinh \rho=-\sqrt{-q}\sin \sigma \ \left(1 +\frac{q}{4} \cos^2
\sigma+O(q^{2})\right)\ , \ \ \ \ \ q\equiv -\sinh^{2} \rho_{0}\ .
\end{equation}
Then the potentials in the fluctuation equations
 are the same as in the $SU(2)$ case in (\ref{pot}).
 The computation of the
 leading  quantum correction to the classical energy
follows  the same steps
 as in the $SU(2)$ case. The only difference is that
  $q=-\sinh^{2} \rho_{0}$ is
now negative. We get as in \rf{kou}
\begin{equation}
E_{1}=\tilde{\lambda}\left[\frac{q}{4}-\frac{11q^2}{128}-
\frac{q^2}{2}\sum_{n=2}^{\infty}\frac{3 n^4+2 n^2+11}{32
(n^2-1)^2}+O(q^3)\right] \ . \la{jop}
\end{equation}
To leading $\tl$ order the expression for  the $AdS_3$  spin $S$
is \cite{ptt}
\begin{equation}
S= J\int_{0}^{2\pi}\frac{d\sigma}{2\pi} \sinh^2 \rho =-\ha q J [ 1
+ \frac{1}{16} q +O(q^2)] \ ,
\end{equation}
where we have given the expansion for the folded solution in the
small $S/J$ limit. Then
\begin{equation}
q=-2\alpha-\ha {\alpha^2}+O(\alpha^3)\ , \ \ \ \ \ \ \ \alpha
\equiv \frac{S}{J}\ .
\end{equation}
Taking into account the expansion  for the
 classical string energy in the short string limit
\cite{ft2}
\begin{equation}
E_{0}=J+S+\frac{\lambda\alpha
}{2J}\left[1-\frac{\alpha}{2}+\frac{3
\alpha^{2}}{8}+O(\alpha^{3})\right]+O(\lambda^2)\ ,
\end{equation}
and adding the regularized expression for $E_1$ in \rf{jop} we
finish with (cf. \rf{hii})
\begin{eqnarray}
E= J&+&S+{\lambda\ov 2 J} \bigg[\a \left(1-\frac{1}{J}+O({ 1 \ov
J^2})\right)\nonumber
\\
&-&
 \ha \a^2 \left(1-\frac{a_2 }{J}+ O({ 1 \ov J^2}) \right)
+  O(\a^3) \bigg]   +  O({\l^2 \ov J^3}) \ ,
\end{eqnarray}
where again
\begin{equation}
a_2=\frac{1}{2}- { \pi^2 \ov 3} \ .
\end{equation}
As in the $SU(2)$ sector, if we formally set $S=2$ we get
\begin{equation}
E=J+2+\frac{\lambda}{J^2}\left(1-\frac{2}{J}+O({1\ov  J^2}) \right) +
...
\end{equation}
which  matches again the near-BMN two-impurity result \ci{callan}.

Note that a similar expression for the energy with the  $1/J$
quantum corrections computed from regularized  quantum LL
Hamiltonian
  can be readily  obtained \ci{btz} in the
case of circular $(S,J)$  solution (for which the full string
1-loop correction was found in
 \cite{ptt}).
 In the small  string  limit one gets
\be E=J+S+{ \lambda  m^2\ov 2 J} \left[ \a \left(1-\frac{1}{J}
+O({ 1 \ov J^2}) \right) + O(\a^2) \right]+ O({\lambda^2\ov J^3})
\ ,
 \ \ \ \ \   \a= {S \ov J} \ .  \ee
%A
%similar expression can be obtained in the case of circular
%solutions in the $SU(2)$ sector.

\end{document}